\documentclass[a4paper, amsfonts, amssymb, amsmath, reprint, showkeys, nofootinbib, twoside,aps,prl]{revtex4-2}
\usepackage[english]{babel}
\usepackage[utf8]{inputenc}
\usepackage{graphicx}
\usepackage{subfigure}
\usepackage{hyperref}
\usepackage{amsmath}
\usepackage{hyphenat}
\usepackage{physics}
\usepackage{amsfonts}
\usepackage{booktabs}
\usepackage{amssymb}
\usepackage[usenames]{color}
\usepackage{lineno}
\usepackage{bm}
\usepackage{mathptmx}

\usepackage{xcolor}
\usepackage{xspace}
\usepackage{lineno}
\usepackage{amssymb}
\usepackage{url}
\usepackage{ulem}
\usepackage{upgreek}
\usepackage{setspace}

\usepackage{graphics}
\usepackage{titlesec}

\newcommand{\figcaption}{\def\@captype{figure}\caption}
\newcommand{\tabcaption}{\def\@captype{table}\caption}

\newcommand{\nineH}        {$\sqrt{s}~=~0.9$~Te\kern-.1emV\xspace}
\newcommand{\seven}        {$\sqrt{s}~=~7$~Te\kern-.1emV\xspace}
\newcommand{\onethree}        {$\sqrt{s}~=~13$~Te\kern-.1emV\xspace}
\newcommand{\twoH}         {$\sqrt{s}~=~0.2$~Te\kern-.1emV\xspace}
\newcommand{\twosevensix}  {$\sqrt{s}~=~2.76$~Te\kern-.1emV\xspace}
\newcommand{\five}         {$\sqrt{s}~=~5.02$~Te\kern-.1emV\xspace}
\newcommand{\twosevensixnn}{$\sqrt{s_{\mathrm{NN}}}~=~2.76$~Te\kern-.1emV\xspace}
\newcommand{\fivenn}       {$\sqrt{s_{\mathrm{NN}}}~=~5.02$~Te\kern-.1emV\xspace}

\newcommand{\GeVc}         {Ge\kern-.1emV/$c$\xspace}
\newcommand{\MeVc}         {Me\kern-.1emV/$c$\xspace}
\newcommand{\GeVmass}      {Ge\kern-.1emV/$c^2$\xspace}
\newcommand{\MeVmass}      {Me\kern-.1emV/$c^2$\xspace}

\newcommand{\lmb}          {\ensuremath{\Lambda}\xspace}

\newcommand{\XiPi}{${\uppi} \Xi$\xspace}

\newcommand{\LKMin}{\ensuremath{K^{-} \Lambda}\xspace}
\newcommand{\JPsiL}{\ensuremath{J/\psi \Lambda}\xspace}

\newcommand{\PiMinXiZ}{\ensuremath{\uppi^- \Xi^0 }\xspace}
\newcommand{\PiZXiMin}{\ensuremath{\uppi^0 \Xi^- }\xspace}
\newcommand{\KMinSigZ}{\ensuremath{\rm K^- \Sigma^0}\xspace}
\newcommand{\KZbarSigMin}{\ensuremath{ \overline{\rm K}^0 \Sigma^-}\xspace}
\newcommand{\EtaXiMin}{\ensuremath{\rm \eta \Xi^- }\xspace}

\newcommand{\XRes}          {\ensuremath{\Xi\mathrm{(1620)}}\xspace}
\newcommand{\XResNovanta}          {\ensuremath{\Xi\mathrm{(1690)}}\xspace}

\begin{document}

\title{Bridging correlation and spectroscopy measurements to access\\ the hadron interaction behind molecular states:\\
the case of the \XRes and \XResNovanta in the \LKMin system}

\author{A. Feijoo}
 \thanks{Corresponding author}
	\email{albert.feijoo@tum.de}
	\affiliation{Physik Department E62, Technische Universit\"at M\"unchen, Garching, Germany, EU}

\author{ V. Mantovani Sarti}
 \email{valentina.mantovani-sarti@tum.de}
	\affiliation{Physik Department E62, Technische Universit\"at M\"unchen, Garching, Germany, EU}
 
\author{J. Nieves} 
 \email{jmnieves@ific.uv.es}
\affiliation{Instituto de F\'{i}sica Corpuscular, Centro Mixto Universidad de Valencia-CSIC, Institutos de Investigaci\'{o}n de Paterna, Aptdo. 22085, E-46071 Valencia, Spain}

\author{A. Ramos}
	\email{ramos@fqa.ub.edu}
	\affiliation{Departament de F\'{i}sica Qu\`antica i Astrof\'{i}sica and Institut de Ci\`encies del Cosmos (ICCUB), Facultat de F\'{i}sica, Universitat de Barcelona, Barcelona, Spain}

  \author{I. Vida\~na}
\email{isaac.vidana@ct.infn.it}
	\affiliation{Istituto Nazionale di Fisica Nucleare, Sezione di Catania, Dipartimento di Fisica ''Ettore Majorana'', Universit\`a di Catania, Via Santa Sofia 64, I-95123 Catania, Italy}

\begin{abstract}
We study the compatibility between the $K^-\Lambda$ correlation function, recently measured by the ALICE collaboration, and the LHCb $K^-\Lambda$ invariant mass distribution obtained in the $\Xi^-_b \to J/\psi \Lambda K^-$ decay. The $K^-\Lambda$ invariant mass distribution associated with the $\Xi^-_b$ decay has been calculated within the framework of Unitary Effective Field Theories using two models, one of them constrained by the $K^-\Lambda$ correlation function. We consider two degenerate pentaquark $P_{cs}$ states in the \JPsiL scattering amplitude which allows us to investigate their impact on both the \LKMin and \JPsiL mass distributions assuming different spin-parity quantum numbers and multiplicity. Without any fitting procedure, the \LKMin model is able to better reproduce the experimental \LKMin mass spectrum in the energy region above $1680$ MeV as compared to previous unitarized scattering amplitudes constrained to a large amount of experimental data in the neutral $S=-1$ meson-baryon sector. We observe a tension between our model and the LHCb \LKMin distribution in the region close to threshold, largely dominated by the presence of the still poorly known \XRes state. We discuss in detail the different production mechanisms probed via femtoscopy and spectroscopy that could provide valid explanations for such disagreement, indicating the necessity to employ future correlation data in other $S=-2$ channels such as \XiPi and $\bar{K}\Sigma$.\\

\end{abstract}

\date{\today}

\maketitle

\section{Introduction}\label{sec:intro} 
The measurement of heavy-hadron weak decays~\cite{LHCb:2024vfz,LHCb:2017iph,LHCb:2019kea,LHCb:2017uwr,LHCb:2020bwg,Belle:2001hyr,Belle:2022voy,Belle:2022ywa,Belle:2024cmc,Belle:2021gtf,Belle:2015wxn,Belle:2011vlx,BaBar:2003oey,BaBar:2004oro,BaBar:2006tck,BaBar:2010wfc,BESIII:2013ouc,ablikim2021study} has become a priceless source of information for characterizing non-fundamental states thanks to the exhaustive analysis carried out by experimental collaborations such as LHCb, Belle, BaBar and BES. Such states appear in the invariant mass spectra of the final decay products and, depending on their nature, they can be either conventional excited states or more exotic structures generated in the same hadronization mechanism or through final state interactions (FSIs). On the theory side, these data have long been considered to infer aspects of the hadron interaction and the formed resonances, as shown in the extensive literature dedicated to these purposes ~\cite{Wu:2024lud,Liu:2023jwo,Li:2023olv,Duan:2024okk,Zhang:2024jby,Ikeno:2024fjr,Abreu:2023rye,Wu:2024lud,Song:2022kac,Wang:2024enc,Miyahara:2016yyh,Feijoo:2015cca,Roca:2015tea,Toledo:2020zxj} (see also the review in~\cite{Oset:2016lyh}).  As a matter of fact, this information is extracted by investigating the scattering processes of the final hadron pairs. However, in contrast to the uniqueness offered by cross-section calculations, there are some inherent difficulties in obtaining informations from the decay processes due to the multiple coexistence of possible mechanisms, and in sorting out their relative weights. Despite the hierarchy of the different weak decay mechanisms is well established, namely the external emission, the internal emission, the $W$ exchange~\cite{Chau:1982da,Chau:1987tk} and, one order of magnitude below, the weak annihilation, the exact theoretical computation of a given decay process remains as a very intricate problem.
One of the recently measured decays is the $\Xi^-_b \to J/\psi \Lambda K^-$ delivered by the LHCb collaboration~\cite{LHCb:2020jpq}. This experimental breakthrough is widely known for reporting the first evidence of a hidden-charm strange pentaquark in the $J/\psi \Lambda$ invariant mass distribution, the so-called $P_{\Psi s}^\Lambda(4459)^0$. Whereas the observation of such exotic state represents a turning point in hadron spectroscopy, the collected data of the $K^-\Lambda$ distribution, and its subsequent analysis, not only allowed an improvement in the precision of the masses and widths for the $\Xi(1690)^-$ and $\Xi(1820)^-$ states but also provided the opportunity to study the meson-baryon interaction in the strangeness $S=-2$ and charge  $Q=-1$ sector.

Having a better knowledge on the $S=-2$ meson-baryon interaction will impact on the understanding of the controversial nature of the $\Xi(1620)$ and $\Xi(1690)$ states (see some discussions in Refs.~\cite{Sekihara:2015qqa,Feijoo:2023wua,Yan:2024usf}). The Unitarized Chiral Perturbation (U$\chi$PT) approach in coupled channels came into play with Ref.~\cite{Ramos:2002xh}, where the dynamically generated state found was assigned to the $\Xi(1620)$ resonance. Slightly after, a similar work~\cite{Garcia-Recio:2003ejq} provided a second pole associated with the $\Xi(1690)$ resonance. The SU$(6)$ extension of the chiral Lagrangian within a coupled channel unitary approach was employed in~\cite{Gamermann:2011mq}, where the $\Xi^*$ poles with $J^P={1/2}^-$ described qualitatively the properties of the $\Xi(1620)$, $\Xi(1690)$ and $\Xi(1950)$ states. In~\cite{Sekihara:2015qqa}, the U$\chi$PT scheme, limited to a contact term, was employed to dynamically generate the $\Xi(1690)$ pole in good agreement with the experimental mass yet with a tiny width, while the $\Xi(1620)$ state was found at an energy more than $50$ MeV below the experimental location. The authors of Ref.~\cite{Khemchandani:2016ftn} incorporated leading-order corrections ($s-, u-,$ and $t-$channel diagrams) to the contact term to obtain a $\Xi(1690)$ pole which was in good agreement with the available experimental data. Very recently, the authors of Ref.~\cite{Nishibuchi:2023acl}, using an approach similar to that of Ref.~\cite{Ramos:2002xh}, pinned the $\Xi(1620)$ down below the $\bar{K}\Lambda$ threshold with a relatively narrow width. Almost simultaneously, in Ref.~\cite{Feijoo:2023wua}, a step forward was made and, for the first time in the present sector, the next-to-leading order (NLO) contributions of the chiral Lagrangian were taken into account. This study showed that the developed model was able to generate dynamically both the $\Xi(1620)$ and the $\Xi(1690)$ states in a very reasonable agreement with the known experimental information and, as in the case of Refs.~\cite{Sekihara:2015qqa,Khemchandani:2016ftn}, the molecular nature of these states naturally explains the puzzle with the decay branching ratios of the $\Xi(1690)$ resonance. It is important to mention that all the previous models were constrained by spectroscopy. More precisely, the parameters present in those models were tuned to reproduce the experimental masses and widths of the $\Xi(1620)$ and $\Xi(1690)$ resonances. 

In the last decade, hadron femtoscopy has been in a good position to prosper as the benchmark technique to provide novel constraints for the hadron interactions, particularly in those sectors where scattering experiments are not feasible. A clear example is the $K^- \Lambda$ correlation function (CF) recently measured by the ALICE collaboration at the LHC~\cite{ALICE:2023wjz}. The analysis of these data performed in~\cite{Sarti:2023wlg} represented the starting point of a novel method to extract information on hadron-hadron scattering amplitudes. More specifically, the low-energy constants (LECs) of the NLO Lagrangian were obtained for the first time using high-precision femtoscopic data (further details of the model and the procedure are given in Sect.~\ref{subsec:FSI}). Moreover, the very recent study of Ref.~\cite{Albaladejo:2024lam} reinforces the robustness and the realistic character of the Koonin–Pratt (KP) formalism~\cite{Koonin:1977fh, Pratt:1990zq,Bauer:1992ffu} to compute CFs. The unequivocal and compact structure of the KP formula leads one to think that the femtoscopy data can provide less ambiguous constraints for the theoretical models compared to those coming from the weak decays.

The present study aims at testing the femtoscopic constraints imposed on the model developed in~\cite{Sarti:2023wlg} by examining its ability to reproduce the experimental $K^-\Lambda$ invariant mass spectrum from $\Xi^-_b \to J/\psi \Lambda K^-$ decay~\cite{LHCb:2020jpq}. In particular, in this work we consider the possibility of different plausible scenarios for 
the $P_{\psi s}^\Lambda$ states within the $J/\psi \Lambda$ invariant mass distribution and we apply significant differences in the modeling of the FSI involving the \LKMin pair.\\
The manuscript is organized in the following way. The theoretical formalism is described in detail in Sec.~\ref{sec:formalism}. Results are presented and discussed in Sec.~\ref{sec:results}. Finally, a short summary and the main conclusions are given in Sec.~\ref{sec:conclusions}.

\section{Formalism}\label{sec:formalism}
\begin{figure}[t]
\begin{center}
\includegraphics[width=0.5\textwidth,keepaspectratio]{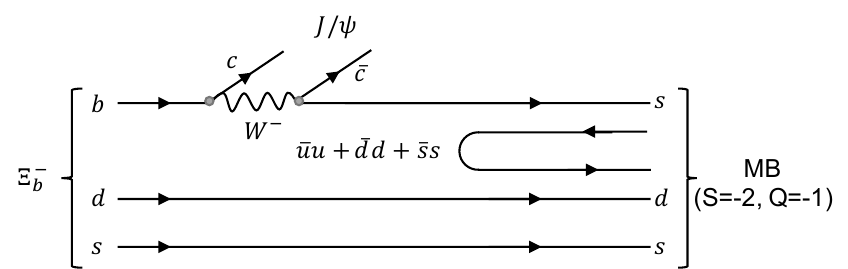}
\caption{(Color online). Internal emission process and hadronization mechanism for the $\Xi^-_b \to J/\psi \Lambda K^-$ decay.}
\label{fig:hadronization}
\end{center}
\end{figure}
In order to describe the $\Xi^-_b \to J/\psi \Lambda K^-$ decay, we follow the same approach as in Ref.~\cite{Chen:2015sxa} by assuming an internal emission weak decay mechanism. 
\subsection{The $\Xi^-_b \to J/\psi \Lambda K^-$ decay process}\label{subsec:decayprocess}
The reaction mechanism, following similar works on different decays~\cite{Roca:2015tea,Feijoo:2015kts,Oset:2016lyh}, consists of a two-step process: a first weak transition and the subsequent hadronization upon which the FSIs take place. Following the discussion of Ref.~\cite{Chen:2015sxa}, we assume that the dominant decay mechanism is the internal emission,
represented diagrammatically at the quark level in Fig.~\ref{fig:hadronization}. 
In the first step, the $d$ and $s$ constituent quarks of the $\Xi_b^-$ act as spectators while the $b$ one decays into a $c$ via the emission of a $W^-$ boson that proceeds through a Cabibbo favored transition into $\bar{c}$ and $s$ quarks. Therefore the initial $\ket{ \Xi_b^-}=\ket{b(ds-sd)}/\sqrt{2}$ flavor wave function is transformed into a $J/\psi$ from the outgoing $\bar{c}c$ pair and a remaining $\ket{s(ds-sd)}/\sqrt{2}$ flavor configuration. The latter undergoes hadronization by incorporating a $\bar{q}q$ pair from the vacuum quantum numbers, $\bar{u}u+\bar{d}d+\bar{s}s$, between the $s$
quark and the spectators, giving rise to the hadronized state
\begin{eqnarray}
\ket{H}&=&\frac{1}{\sqrt{2}}\ket{s(\bar{u}u+\bar{d}d+\bar{s}s)(ds-sd)} \nonumber \\ 
&=&\frac{1}{\sqrt{2}}\sum_{i=1}^3P_{3i}\,q_i(ds-sd) \, ,
\label{hadronization1}    
\end{eqnarray}
where in the second line we have introduced the
SU(3) flavor matrix arrangement
\begin{equation}\label{eq:PandPhimatrices}
P=\left(\begin{array}{ccc}
u\bar{u} & u\bar{d} & u\bar{s}\\
d\bar{u} & d\bar{d} & d\bar{s}\\
s\bar{u} & s\bar{d} & s\bar{s}
\end{array}\right) \ .
\end{equation}
Replacing it by the pseudoscalar-meson fields matrix
\begin{equation}
\phi=\left(\begin{array}{ccc}\frac{\pi^0}{\sqrt{2}}+\frac{\eta}{\sqrt{3}}+\frac{\eta'}{\sqrt{6}} & \pi^+ & K^+\\
\pi^- & -\frac{\pi^0}{\sqrt{2}}+\frac{\eta}{\sqrt{3}}+\frac{\eta'}{\sqrt{6}} & K^0\\
K^- & \bar{K}^0 & -\frac{\eta}{\sqrt{3}}+\frac{2\eta'}{\sqrt{6}}\end{array}\right)\, \, ,
\end{equation}

the hadronized state can be written as
\begin{eqnarray}
\ket{H}&=&\frac{1}{\sqrt{2}}\sum_{i=1}^3\phi_{3i}\, q_i(ds-sd)= K^- \Big[\frac{1}{\sqrt{2}}u(ds-sd)\Big] \nonumber \\ 
& + &  \bar{K}^0\Big[\frac{1}{\sqrt{2}}d(ds-sd)\Big]-\frac{\eta}{\sqrt{3}}\Big[\frac{1}{\sqrt{2}}s(ds-sd)\Big] \nonumber \\ 
&=& \frac{1}{\sqrt{2}}K^-\Sigma^0 + \bar{K}^0 \Sigma^- - \frac{1}{\sqrt{6}}K^-\Lambda -\frac{1}{\sqrt{3}} \eta \Xi^- ,  
\label{hadronization2}    
\end{eqnarray}
where we assume the standard $\eta-\eta'$ mixing, and we neglect the $\eta'$ meson contribution due to its heavy mass. It should be noted that in the last row of Eq.~(\ref{hadronization2}), we rewrite the mixed antisymmetrized representation of the ground-state baryons from the remaining three-quark combinations following Ref.~\cite{Close1979}. It is then clear that the hadronization process ends up producing several $S=-2$ meson-baryon (MB) pairs and their FSIs must be accounted for in the calculation of the \LKMin invariant mass spectrum.

\begin{figure}[h!]
\begin{center}
\includegraphics[width=0.495\textwidth,keepaspectratio]{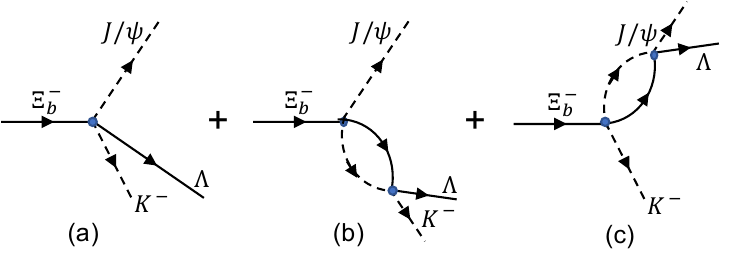}
\caption{(Color online). Feynman diagrams considered in the evaluation of the $\Xi^-_b \to J/\psi \Lambda K^-$ decay amplitude (see text for details).} 
\label{fig:feynmandiag}
\end{center}
\end{figure}
\begin{figure*}[t]
\begin{center}
\includegraphics[width=0.8\textwidth,keepaspectratio]{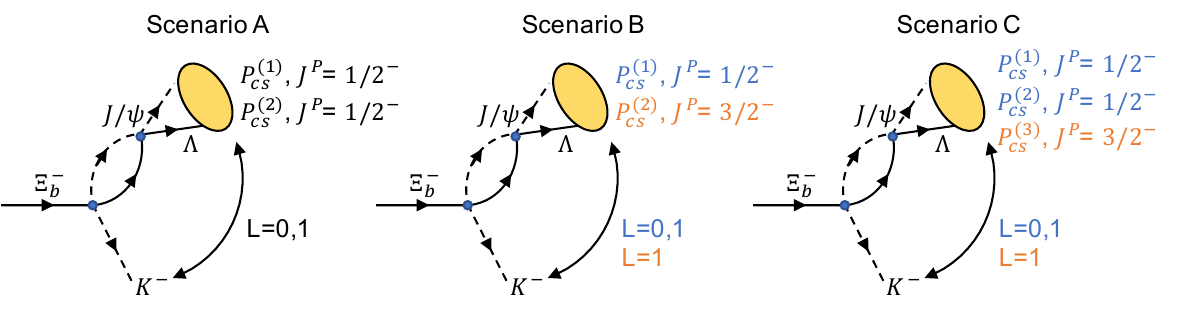}
\caption{(Color online). Three scenarios assumed in the model for the final pentaquark states.}
\label{fig:scenariosPcs}
\end{center}
\end{figure*}

Hence, apart from the primary hadron production accounted for by the tree level diagram in Fig.~\ref{fig:feynmandiag} (diagram~(a)), one should also consider contributions coming from the intermediate ($S=-2,Q=-1$) MB pairs whose interaction leads to the final $K^- \Lambda$ (diagram (b)) in Fig.~\ref{fig:feynmandiag}), but also the contribution from diagram (c) that accounts for the $J/\psi \Lambda \to J/\psi \Lambda$ FSI. All these terms enter in the following expression for the amplitude:
\begin{eqnarray}
& &\mathcal{M}(m_{K^- \Lambda},m_{J/\psi \Lambda}) = V_{p} \Big[ h_{K^-\Lambda}  
+  \sum_i h_i G_i( m_{K^-\Lambda}) t_{i,K^-\Lambda}( m_{K^-\Lambda}) \nonumber \\
&  &+ \, h_{K^-\Lambda} G_{J/\psi \Lambda}(m_{J/\psi \Lambda}) \, t_{J/\psi \Lambda,J/\psi \Lambda}(m_{J/\psi \Lambda} )\Big] ,
\label{eq:amplitude1}
\end{eqnarray}
where the quantities $h_i$ ($i=\pi^- \Xi^0$, $\pi^0 \Xi^-$, $K^- \Lambda$, $K^- \Sigma^0$, $\bar{K}^0 \Sigma^-$, $\eta \Xi^- $) are the transition weights, which can be directly obtained from Eq.~(\ref{hadronization2}),
\begin{eqnarray}
 \label{h_weights}   
 h_{\pi^- \Xi^0}=0 \,,\,\,  & h_{\pi^0 \Xi^-}=0 \, , & \,\, h_{K^- \Lambda}=-\frac{1}{\sqrt{6}}\, , \nonumber \\
  h_{K^- \Sigma^0}=\frac{1}{\sqrt{2}} \,,  \,\,& h_{\bar{K}^0 \Sigma^-}=1 \, , &\,\, h_{\eta \Xi^- }=-\frac{1}{\sqrt{3}} \,.  
\end{eqnarray} 
The quantity $G_i$ ($G_{J/\psi \Lambda}$) stands for the $i$ ($\JPsiL$) MB loop,
defined in accordance with the theoretical models employed for the $t_{i,K^-\Lambda}$ ($t_{J/\psi \Lambda,J/\psi \Lambda}$) scattering amplitudes (see Sect.~\ref{subsec:FSI} below for a complete description). In addition, $m_{K^-\Lambda}$ and $m_{J/\psi \Lambda}$ are the invariant masses of the $K^-\Lambda$ and $J/\psi \Lambda$ pairs respectively. The factor $V_{p}$ represents the strength of the weak process occurring at the quark level (see Fig.~\ref{fig:hadronization}), whose value is a-priori unknown and is typically assumed to be constant (the argumentation can be found in Ref.~\cite{Feijoo:2015cca}). The amplitude in Eq.~(\ref{eq:amplitude1}) is written as a coherent sum of terms containing the scattering amplitudes $t_{i,K^-\Lambda}$ and $t_{J/\psi \Lambda,J/\psi \Lambda}$, obtained within a unitarized coupled-channel approach (for more details see Sec.~\ref{subsec:FSI}). Both amplitudes are assumed to be in s-wave, hence lacking any angular or spin dependence. This assumption is justified by the fact that, when using the unitary extension of such chirally motivated interactions, the strength of the s-wave contribution close to threshold is much larger than the p-wave ones.
However, despite the $J/\psi \Lambda$ interaction is considered to be in s-wave, the intrinsic spin-parity of the $J/\psi$ ($J^P _{J/\psi}=1^-$) and $\Lambda$ ($J^P _{\Lambda}=1/2^+$) hadrons leads to total $J^P _{J/\psi \Lambda} =1/2^-$ or $3/2^-$. In the amplitude $\mathcal{M}$ describing the weak $\Xi^-_b \to J/\psi \Lambda K^-$ decay, the total angular momentum of the initial $\Xi^-_b$ state ($J^P _{\Xi^-_b}= 1/2^+$) has to be conserved, and since the $K^-$ carries zero spin ($J^P _{K^-}=0^-$), further considerations must be taken into account when $J^P _{J/\psi \Lambda} = 3/2^-$.\\
More precisely, for the case where $J/\psi \Lambda$ and $K^-\Lambda$ have spin-parity equal to $1/2^-$ and both systems are in s-wave with respect to the remaining meson ($K^-$ or $J/\psi$ respectively), one can guarantee the conservation of the initial $J=1/2$ from the $\Xi_b^-$.  This is not the case when the $J/\psi \Lambda$ is in $J^P=3/2^-$ and a $L=0$ relative angular momentum of the $K^-$ with respect to $J/\psi \Lambda$ pair is considered. As discussed in Refs.~\cite{Lu:2016roh,Chen:2015sxa}, for the case $J^P _{J/\psi \Lambda} = 3/2^-$ only a p-wave production vertex with the $K^-$ allows the matching with the $\Xi_b^-$ angular momentum.
The full squared amplitude accounting for all these considerations in the $\Xi^-_b \to J/\psi \Lambda K^-$ decay is given by~\cite{Lu:2016roh,Chen:2015sxa}
\begin{eqnarray}
|\mathcal{M}|^2 &=& \mathcal{N}\Big [3V^2_p  \Big| T_{1/2} ^{\rm{s-wave}}\Big|^2+\frac{3}{2}\vec{q}^2 B^2 \Big|T_{3/2} ^{\rm{p-wave}}\Big|^2
\nonumber \\ 
&+& 3\vec{q}^2 B'^2 \Big| T_{1/2} ^{\rm{p-wave}}\Big|^2
+ \sum_i |t_{\Xi^{\ast} _i} ^{BW}(m_{\LKMin})|^2
\Big]\,,
\label{eq:Genamplitude2}
\end{eqnarray}
where a sum over the initial polarizations and an average over the final ones have been performed. The values of  $\mathcal{N}$, $V_p$, $B$ and $B'$, as it is explained later (c.f. Sec.~\ref{subsec:normalization}), are estimated either by normalizing the theoretical amplitude to the LHCb data \cite{LHCb:2020jpq} or via theoretical considerations. $T_{J_{J/\psi \Lambda}}^{L}$ represents the total amplitude given by the sum of the Feynman diagrams of Fig.~\ref{fig:feynmandiag} assuming a specific spin $J_{J/\psi \Lambda}$ for the $J/\psi \Lambda$ system and a relative angular momentum $L$ between it and the $K^-$. Note that when one considers only $J_{J/\psi \Lambda}=1/2$ and $L=0$, 
Eq.~(\ref{eq:Genamplitude2}) reduces essentially to the modulus square of Eq.~(\ref{eq:amplitude1}). The last term in Eq.~(\ref{eq:Genamplitude2}), which includes the contribution of the three resonances ($i=\Xi\mathrm{(1820)},\Xi\mathrm{(1950)}, \Xi\mathrm{(2030)}$) observed in the \LKMin invariant mass data, is added incoherently in the same spirit of the LHCb experimental analysis of Ref.~\cite{LHCb:2020jpq}.\\
The transition amplitude accounting for the $\Xi$(1820) can be parametrized as
\begin{align}\label{eq:BWPheno}
t_{\Xi(1820) }^{BW}(m_{\LKMin}) = \dfrac{g_{\LKMin} ^2}{m_{\LKMin} - M_{\Xi(1820)}+i\frac{\Gamma_{\Xi(1820)}}{2}}
\end{align}
with the mass and the width of the resonant state taken from Table 2 of Ref.\ \cite{LHCb:2020jpq}, and the coupling $g_{\LKMin}$  fixed to reproduce the branching ratio to \LKMin reported in the Particle Data group~\cite{PDG} ($\Gamma(\bar{K}\Lambda)/\Gamma_{\rm total}=0.25$ with $\Gamma_{\rm total}\approx 8$~MeV~\cite{LHCb:2020jpq}). More specifically, $g_{\LKMin}$ was obtained following Eq.~(C15) in appendix C of Ref.~\cite{Penner:2002ma} particularized for the case of a $J^P=3/2^-$ state ($\Xi(1820)$) decaying into a meson ($K^-$) with identical parity. The contribution of the other two resonances, $\Xi$(1950) and $\Xi$(2030), is modelled using a relativistic Breit--Wigner (BW) given in Eq.~(2) of Ref.\ \cite{Giacosa:2021mbz}, with masses and widths taken from the values reported in the LHCb fit (see Table 2 in ~\cite{LHCb:2020jpq}).\\
Let us focus on the contribution to the final amplitude of diagram (c) in Fig.~\ref{fig:feynmandiag} of the pentaquark states dynamically generated by the $J/\psi \Lambda \to J/\psi \Lambda$ FSI. In the present work, we analyze three different scenarios A, B and C (see  Fig.~\ref{fig:scenariosPcs}) depending on the assumed value of the total spin-parity 
$J^P_{P_{cs}}$ ($1/2^-$ or $3/2^-$) of the pentaquark states, and on the relative angular momentum $L$ between the $K^-$ and the $J/\psi \Lambda$ system. Particularly, in scenario A 
we consider two pentaquark states both with spin-parity $J^P_{P_{cs}} = 1/2^-$. The lower-mass one is identified with the state $P_{\Psi s}^\Lambda(4338)^0$, in line with the experimental analysis carried out in \cite{LHCb:Pcs_4338_2022} while the higher-mass one is associated with $P_{\Psi s}^\Lambda(4459)^0$~\cite{LHCb:Pcs_4459_2020}. In scenario B, instead, we assume a different spin-parity for the two pentaquaks: $J^P_{P_{cs}} = 1/2^-$ for the lower-mass one and $J^P_{P_{cs}} = 3/2^-$ for the higher-mass state.
Finally, by analogy with the observed non-strange pentaquarks $P_\Psi^N(4312), P_\Psi^N(4440)$ and $P_\Psi^N(4457)$~\cite{LHCb:2019kea}, in scenario C we also explore the possibility of having a third pentaquark with $J^P_{P_{cs}} = 3/2^-$ hidden in the structure of the apparently single $P_{\Psi s}^\Lambda(4459)$ state (assumed to have $J^P_{P_{cs}} = 1/2^-$) or in the fluctuations arising at slightly higher energy in the LHCb data. In the following, we describe in detail the three scenarios assumed for the pentaquark states, and how the corresponding total amplitudes $T_{J_{J/\psi \Lambda}}^{L}$ entering in Eq.\ (\ref{eq:Genamplitude2}) read in each particular case.
\vspace{0.5em}

\subsubsection*{Scenario A: two $P_{cs}$ states with $J^P_{P_{cs}} = 1/2^-$}
In this case, we consider a low and a high mass pentaquark state, each one with spin-parity $1/2^-$. As explained before, the two states  can contribute to both s- and p-wave production vertices with respect to the  $K^-$. The amplitude terms entering in Eq.~\ref{eq:Genamplitude2} are explicitly defined as
\begin{eqnarray}\label{eq:AmplscenarioA}
    T_{1/2} ^{\rm{s-wave}} & = & h_{\LKMin} + \nonumber \\
    & &+\sum_i h_i G_i( m_{K^-\Lambda}) t_{i,K^-\Lambda}( m_{K^-\Lambda}) \nonumber \\
   & &+ h_{\LKMin} G_{J/\psi \Lambda}(m_{J/\psi \Lambda}) \, t_{J/\psi \Lambda,J/\psi \Lambda}(m_{J/\psi \Lambda} )\, ,\nonumber\\
    T_{3/2} ^{\rm{p-wave}} & = & h_{\LKMin} \, ,\nonumber\\
    T_{1/2} ^{\rm{p-wave}} & = & h_{\LKMin} +\nonumber \\
    & &+ h_{\LKMin} G_{J/\psi \Lambda}(m_{J/\psi \Lambda}) \, t_{J/\psi \Lambda,J/\psi \Lambda}(m_{J/\psi \Lambda} )\, .
\end{eqnarray}
We consider the $t_{J/\psi \Lambda,J/\psi \Lambda}$ scattering amplitude given in Ref.~\cite{Feijoo:2022rxf} (labeled as Valencia (VLC) model) which delivers two poles in the vector meson-baryon ($J^P=1/2^-$) sector, each with two degenerate states having $J^P = 1/2^-,3/2^-$. Since in this scenario no pentaquark is assumed to have spin $3/2$, only the tree-level amplitude is considered in $T_{3/2}^{\rm{p-wave}}$. In the $T_{1/2} ^{\rm{s-wave}}$ term, we also have to take into account the $\LKMin$ contribution. For the latter, we employ the meson-baryon $S=-2$ scattering amplitude obtained within a unitarized chiral effective lagrangian up to NLO. In particular, we will show the comparison to the \LKMin invariant mass spectrum measured by the LHCb collaboration assuming the parametrization obtained in~\cite{Feijoo:2023wua} (Barcelona (BCN) model) and the one constrained to the measured $\LKMin$ correlation function~\cite{Sarti:2023wlg} (Valencia-Barcelona-Catania (VBC) model). More details on the modeling of the $J/\psi \Lambda$ and \LKMin interactions are discussed in Sect.~\ref{subsec:FSI}.
\vspace{0.5em}
\subsubsection*{Scenario B: two $P_{cs}$ states with $J^P_{P_{cs} ^{(1)}} = 1/2^-$ and $J^P_{P_{cs} ^{(2)}} = 3/2^-$}
\vspace{0.5em}
In this case we assume the high mass $P_{cs} ^{(2)}$ pentaquark state to have spin $3/2$, hence requiring a p-wave vertex with the $K^-$ to conserve the total angular momentum. Thus the two pentaquarks have different $J$ and, therefore, the use of the $t_{J/\psi \Lambda,J/\psi \Lambda}$ from VLC model, which simultaneously accounts for the contribution of both states, should be avoided to prevent double counting. Instead, we follow the same recipe used for the $\Xi$(1820) and assume a phenomenological Breit--Wigner amplitude $t_{P_{cs} ^{(i)}} ^{BW}(m_{J/\psi \Lambda})$ for each pentaquark state ($i=1, 2$), with the couplings $g_{i\,J/\psi \Lambda}$ to the \JPsiL channel given in Table~\ref{tab:spectroscopy} (see Sect.~\ref{subsec:FSI} for more details).
Based on this, the corresponding amplitudes in Eq.~(\ref{eq:Genamplitude2}) read
\begin{eqnarray}\label{eq:AmplscenarioB}
    T_{1/2} ^{\rm{s-wave}} & = & h_{\LKMin} + \nonumber \\
    & &+\sum_i h_i G_i( m_{K^-\Lambda}) t_{i,K^-\Lambda}( m_{K^-\Lambda}) \nonumber \\
   & &+ h_{\LKMin} G_{J/\psi \Lambda}(m_{J/\psi \Lambda}) \, t_{P_{cs} ^{(1)}} ^{BW}(m_{J/\psi \Lambda} )\, ,\nonumber\\
    T_{3/2} ^{\rm{p-wave}} & = & h_{\LKMin} + \nonumber \\
    & & + h_{\LKMin} G_{J/\psi \Lambda}(m_{J/\psi \Lambda}) \, t_{P_{cs} ^{(2)}} ^{BW}(m_{J/\psi \Lambda} )\, ,\nonumber\\
    T_{1/2} ^{\rm{p-wave}} & = & h_{\LKMin} +\nonumber \\
    & &+ h_{\LKMin} G_{J/\psi \Lambda}(m_{J/\psi \Lambda}) \, t_{P_{cs} ^{(1)}} ^{BW}(m_{J/\psi \Lambda} ).
\end{eqnarray}
\vspace{0.5em}
\raggedbottom
\subsubsection*{Scenario C: three $P_{cs}$ states with $J^P_{P_{cs} ^{(1)}} = 1/2^-$, $J^P_{P_{cs} ^{(2)}} = 1/2^-$, $J^P_{P_{cs} ^{(3)}} = 3/2^-$}
\vspace{0.5em}
In this final case, we assume a spin $1/2$ low mass $P_{cs} ^{(1)}$ pentaquark and a degeneracy in spin for the high mass ones which leads to having a second $P_{cs} ^{(2)}$ and a third $P_{cs} ^{(3)}$ state with spin 1/2 and 3/2, respectively. Under these assumptions, the \JPsiL amplitude $t_{J/\psi \Lambda,J/\psi \Lambda}$ already takes into account the contributions from the $P_{cs} ^{(1)}$ and $P_{cs} ^{(2)}$ states, while 
to include the contribution of the $P_{cs} ^{(3)}$ pentaquark we employ, similarly to scenario B, a Breit--Wigner type amplitude.
The final amplitudes hence become
\begin{eqnarray}\label{eq:AmplscenarioC}
    T_{1/2} ^{\rm{s-wave}} & = & h_{\LKMin} + \nonumber \\
    & &+\sum_i h_i G_i( m_{K^-\Lambda}) t_{i,K^-\Lambda}( m_{K^-\Lambda}) \nonumber \\
     & &+ h_{\LKMin} G_{J/\psi \Lambda}(m_{J/\psi \Lambda}) \, t_{J/\psi \Lambda,J/\psi \Lambda}(m_{J/\psi \Lambda} )\, ,\nonumber
\end{eqnarray}
\begin{eqnarray}\label{eq:AmplscenarioCbis}
    T_{3/2} ^{\rm{p-wave}} & = & h_{\LKMin} + \nonumber \\
    & & + h_{\LKMin} G_{J/\psi \Lambda}(m_{J/\psi \Lambda}) \, t_{P_{cs} ^{(3)}} ^{BW}(m_{J/\psi \Lambda} )\, ,\nonumber \\
    T_{1/2} ^{\rm{p-wave}} & = & h_{\LKMin} +\nonumber \\
    & &+ h_{\LKMin} G_{J/\psi \Lambda}(m_{J/\psi \Lambda}) \, t_{J/\psi \Lambda,J/\psi \Lambda}(m_{J/\psi \Lambda} ).
\end{eqnarray}

\subsection{Theoretical models for the FSI} 
\label{subsec:FSI}
\begin{figure*}[t]
\begin{center}
\includegraphics[width=0.75\textwidth,keepaspectratio]{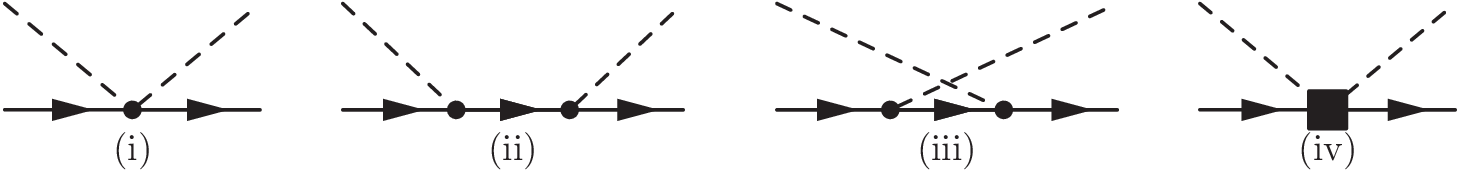}
\caption{Feynman diagrams for meson-baryon interaction: WT term (i), direct and crossed Born terms (ii) and (iii), and NLO terms (iv). Dashed (solid) lines represent the pseudoscalar octet mesons (octet baryons).}
\label{fig:diagrams}
\end{center}
\end{figure*}
We describe now in detail the theoretical approach employed to compute the FSIs taking place in diagrams (b) and (c) of Fig.~\ref{fig:feynmandiag}. The hadron pairs involved in each diagram belong to different sectors, whose interaction is derived from a unitary extensions of different Lagrangians. In the case of diagram (b), we deal with the interaction between a pseudoscalar meson ($\phi$) and a $J^P=1/2^-$ ground state baryon (B) in the $S=-2$ sector, while the FSI in diagram (c) involves a vector meson (V) and a $J^P=1/2^-$ ground state baryon (B) in the hidden charm ($\bar{c}c$) $S=-1$ sector.\\

The starting point to compute the $\phi$B interaction is the $SU(3)$ chiral effective Lagrangian up to NLO $\mathcal{L}_{\phi B}^{eff}=\mathcal{L}_{\phi B}^{(1)}+\mathcal{L}_{\phi B}^{(2)}$, which reads
\begin{eqnarray} 
\mathcal{L}_{\phi B}^{(1)} & = & i \langle \bar{B} \gamma_{\mu} [D^{\mu},B] \rangle
                            - M_0 \langle \bar{B}B \rangle  
                           + \frac{1}{2} D \langle \bar{B} \gamma_{\mu} 
                             \gamma_5 \{u^{\mu},B\} \rangle \nonumber \\
                  & &      + \frac{1}{2} F \langle \bar{B} \gamma_{\mu} 
                               \gamma_5 [u^{\mu},B] \rangle \ ,
\label{LagrphiB1} \\
  \mathcal{L}_{\phi B}^{(2)}& = & b_D \langle \bar{B} \{\chi_+,B\} \rangle
                             + b_F \langle \bar{B} [\chi_+,B] \rangle
                             + b_0 \langle \bar{B} B \rangle \langle \chi_+ \rangle  \nonumber \\ 
                     &  & + d_1 \langle \bar{B} \{u_{\mu},[u^{\mu},B]\} \rangle 
                            + d_2 \langle \bar{B} [u_{\mu},[u^{\mu},B]] \rangle     \nonumber \\
                    &  &  + d_3 \langle \bar{B} u_{\mu} \rangle \langle u^{\mu} B \rangle
                            + d_4 \langle \bar{B} B \rangle \langle u^{\mu} u_{\mu} \rangle \ .
\label{LagrphiB2}
\end{eqnarray}
The $\mathcal{L}_{\phi B}^{(1)}$ contains the contact term, corresponding to the Weinberg--Tomozawa (WT) contribution (diagram (i) in Fig.~\ref{fig:diagrams}), and the constituents of the vertices employed to construct direct and crossed Born terms schematically represented by diagrams (ii) and (iii), whereas the tree-level NLO contributions are fully extracted from $\mathcal{L}_{\phi B}^{(2)}$ (diagram (iv) in Fig.~\ref{fig:diagrams}). Regarding the inputs entering in Eqs.~(\ref{LagrphiB1}) and (\ref{LagrphiB2}), $B$ is the octet baryon matrix, while the matrix of the pseudoscalar mesons $\phi$  is implicitly contained in $u_\mu = i u^\dagger \partial_\mu U u^\dagger$, where $U(\phi) = u^2(\phi) = \exp\left( \sqrt{2} {\rm i} \phi/f \right) $ with $f$ being the effective meson decay constant. The covariant derivative is given by  $[D_\mu, B] = \partial_\mu B + [ \Gamma_\mu, B]$, with $\Gamma_\mu =  [ u^\dagger,  \partial_\mu u] /2$, while $\chi_+ = 2 B_0 (u^\dagger \mathcal{M} u^\dagger + u \mathcal{M})$, with $\mathcal{M} = {\rm diag}(m_u, m_d, m_s)$ and $B_0 = - \langle 0 |\bar{q} q | 0 \rangle / f^2$, is the explicit chiral symmetry breaking term. In addition, $D$ and $F$ are the axial vector constants and $M_0$ stands for the baryon octet mass in the chiral limit. However, as usually done in this type of effective Lagrangian approaches, we employ the physical baryon (and meson) masses in the computation of the interaction. The $b_D$, $b_F$, $b_0$ and $d_i$ $(i=1,\dots,4)$ coefficients preceding the NLO terms considered in Eq.~(\ref{LagrphiB2}) are low energy constants (LECs) to be determined from experimental measurements. At the end, the total interaction kernel derived from Eqs.~(\ref{LagrphiB1}) and (\ref{LagrphiB2}) can be expressed as $V_{ij}=V^{\scriptscriptstyle WT}_{ij}+V^{D}_{ij}+V^{C}_{ij}+V^{\scriptscriptstyle NLO}_{ij}$, where the analytical forms of the different pieces entering in the interaction matrix ${V}_{ij}$ can be found in Refs.~\cite{Borasoy:2005ie,Hyodo:2011ur,Ramos:2016odk,Feijoo:2021zau}.

Since the decay amplitude relies on the scattering amplitude $t_{ij}$, as final step, we solve the Bethe--Salpeter (BS) equation through an on-shell factorization, leaving a simple system of algebraic equations expressed in matrix form as
\begin{equation}
t_{ij} ={(1-V_{il}G_l)}^{-1}V_{lj} ,
 \label{T_algebraic}
\end{equation}
 with $G_l$ being the meson-baryon loop function whose logarithmic divergence is handled by dimensional regularization
\begin{eqnarray}
 G_l & = &\frac{2M_l}{(4\pi)^2} \Bigg \lbrace a_l(\mu)+\ln\frac{M_l^2}{\mu^2}+\frac{m_l^2-M_l^2+s}{2s}\ln\frac{m_l^2}{M_l^2}  \nonumber \\ 
 &  +   &  \frac{q_{\rm cm}}{\sqrt{s}}\ln\left[\frac{(s+2\sqrt{s}q_{\rm cm})^2-(M_l^2-m_l^2)^2}{(s-2\sqrt{s}q_{\rm cm})^2-(M_l^2-m_l^2)^2}\right]\Bigg \rbrace.  
 \label{dim_reg}    
\end{eqnarray}
Note that this latter expression comes in terms of the baryon $M_l$ (meson $m_l$) masses for the $l$-channel as well as the so-called subtraction constants (SCs) $a_l$, which replace the divergence for a given dimensional regularization scale $\mu$ ($\mu=630$~MeV in the present case). In principle, there are as many SCs as channels considered in the sector at hand, but this number can be reduced applying isospin symmetry arguments. Furthermore, and despite one can assign a natural size to all $a_l$'s (see Ref.~\cite{Oller:2000fj}), it is common practice in these kind of approaches to include them in the fitting procedure, within a reasonable range, in order to provide more versatility to the models.

For the present study, this formalism should be applied to the ($S=-2$, $Q=-1$) sector. Altogether, the U$\chi$PT with WT+Born+NLO terms in this sector leaves a scattering amplitude that depends on $14$ parameters. As far as we know, the $S=-2$ $\phi$B interaction extended to NLO has only been addressed in Refs.~\cite{Feijoo:2023wua} and \cite{Sarti:2023wlg} employing the BCN model and the VBC one, respectively. In BCN model, most of the parameters ($D$, $F$ and the NLO LECs) were assumed to be $SU(3)$ symmetric and taken from those of Ref.~\cite{Feijoo:2018den}, where they were fitted to the large amount of experimental data in the neutral sector $S=-1$. In addition, the effective decay constant $f$ and the $4$ $a_l$ SCs were tuned to simultaneously reproduce the properties of the \XRes and \XResNovanta states in the best possible way according to their experimental values \cite{LHCb:2020jpq,BelleXi}. 

On the other hand, the authors of \cite{Sarti:2023wlg} presented a novel method to extract information on hadron-hadron interactions using high-precision femtoscopic data to constrain the low-energy constants of a QCD effective Lagrangian. This study focused on the $K^-\Lambda$ CF and, as a result, all the LECs of the effective Lagrangian up to NLO in the $S=-2$ sector were determined for the first time (see a detailed explanation in \cite{Sarti:2023wlg}). The model obtained from this procedure and used in the present study is referred to as the VBC model.
\begin{table}[t!]
\centering
\caption{Poles of the resonances generated by the VBC and BCN $S=-2$ meson-baryon interaction at NLO corresponding to the \XRes and \XResNovanta states.}
\begin{tabular}{l|cc|cc}
\multicolumn{5}{c}{}\\[-2.5mm]
\hline
\hline \multicolumn{5}{c}{}\\[-3.5mm]
 \multicolumn{1}{c}{}  &  \multicolumn{2}{c}{\XRes}  &  \multicolumn{2}{c}{\XResNovanta}   \\
 \hline
  \multicolumn{1}{c}{Model VBC~\cite{Sarti:2023wlg}}  &  \multicolumn{2}{c}{}  &  \multicolumn{2}{c}{}   \\
 \multicolumn{1}{c}{mass $M$ [MeV]:}  &  \multicolumn{2}{c}{$1612.68$}  &  \multicolumn{2}{c}{$1670.28$}   \\
 \multicolumn{1}{c}{width $\Gamma$ [MeV]:}   &       \multicolumn{2}{c}{$24.57$} &      \multicolumn{2}{c}{$7.44$} \\ 
  \hline
  \multicolumn{1}{c}{Model BCN~\cite{Feijoo:2023wua}}  &  \multicolumn{2}{c}{}  &  \multicolumn{2}{c}{}   \\
 \multicolumn{1}{c}{mass $M$ [MeV]:}  &  \multicolumn{2}{c}{$1606.86$}  &  \multicolumn{2}{c}{$1680.99$}   \\
 \multicolumn{1}{c}{width $\Gamma$ [MeV]:}   &       \multicolumn{2}{c}{$171.18$} &      \multicolumn{2}{c}{$5.17$} \\ 
\hline
 \multicolumn{1}{c}{Experimental $\Xi^*$:}     &      \multicolumn{2}{c}{\XRes~\cite{BelleXi}} & \multicolumn{2}{c}{\XResNovanta~\cite{PDG}} \\
 \multicolumn{1}{c}{mass $M$ [MeV]:}  &  \multicolumn{2}{c}{$1610.4\pm6.0^{+5.9}_{-3.5}$}  &  \multicolumn{2}{c}{$1690\pm10$}   \\
 \multicolumn{1}{c}{width $\Gamma$ [MeV]:}   &       \multicolumn{2}{c}{$59.9\pm4.8^{+2.8}_{-3.0}$} &      \multicolumn{2}{c}{$20\pm15$} \\ 
\hline
 \multicolumn{5}{c}{}\\[-4.2mm]
\hline
\end{tabular}
\label{tab:poles}
\end{table}

To aid the discussion of the results in Sect.~\ref{sec:results}, we compile in Table~\ref{tab:poles} the pole properties (mass and width) for the \XRes and \XResNovanta states obtained within the VBC and BCN models. In both schemes, the corresponding masses and widths for the two poles are compatible with the current experimental data~\cite{PDG,BelleXi}, reported in the last two rows. The agreement with the experimental data is an evidence that the inclusion of Born and the NLO contributions plays a key role to dynamically generate the \XRes and \XResNovanta resonances simultaneously.

The main difference between these two models comes when inspecting the couplings of the different channels to the \XRes pole. By comparing Table~3 in \cite{Feijoo:2018den} to Table~II in \cite{Sarti:2023wlg}, one immediately notes the change of paradigm for this resonance. The strong coupling to the $\eta \Xi$ channel, along with a factor $2$ reduction of the coupling to the $\bar{K} \Lambda$ found in the VBC model, are in contrast to what was obtained in the BCN model, {\it i. e.,} a \XRes state couples mostly to $\pi \Xi$ and $\bar{K} \Lambda$ channels. The fact that, in the VBC model, this state is basically a $\eta \Xi$ molecule will have a direct consequence in the decay amplitude. The \XResNovanta resonance appears basically as a $\bar{K} \Lambda$ quasi-bound state in both models. It can also be observed that the theoretical energy location is below the expected experimental value in both models.\\

In order to address the VB interaction taking place in diagram (c) of Fig.~\ref{fig:feynmandiag}, we resort to the formalism of Ref.~\cite{Feijoo:2022rxf} that naturally evolved from the original work of \cite{Xiao:2019gjd}. This revisited model was designed for different families in this sector, namely pseudoscalar meson-baryon of $J^P=1/2^+$, pseudoscalar meson-baryon of $J^P=3/2^+$, vector meson-baryon of $J^P=1/2^+$ and vector meson-baryon of $J^P=3/2^+$. We limit the present study to the VB ($1/2^+$) family that incorporates the following channels: $J/\psi\Lambda(4212.6)$, $\bar{D}^*_s\Lambda_c(4398.7)$, $\bar{D}^*\Xi_c(4477.6)$, $\bar{D}^*\Xi_c^\prime(4587.0)$. This approach is based on heavy quark spin symmetry (HQSS) together with the local hidden gauge approach (LHG) \cite{Bando:1984ej,Bando:1987br,Meissner:1987ge} extrapolated to the charm sector. The interaction derived from the LHG consists of a $t-$channel diagram whose mediating particle is a vector meson. As it is well known, this diagram can be reduced to a contact term when the transferred momentum to the exchanged gauge vector meson is much smaller than its mass. In summary, the transition potentials between the channels considered in s-wave read 
\begin{eqnarray}
  \label{eq:def_Vij}
     V_{ij}= C_{ij}\frac{1}{4f_\pi^2}(p^0+p^{\prime\,0})\, ,
\end{eqnarray}
where $f_\pi=93$ MeV, and $p^0, p^{\prime\,0}$ are the energies of the incoming and outgoing mesons involved in the corresponding transitions. Since the decay mechanism depicted in Fig.~\ref{fig:hadronization} primarily produces the $J/\psi$ particle, the only intermediate channel that can be taken into account in diagram (c) of Fig.~\ref{fig:feynmandiag} is the $J/\psi \Lambda$ one. Thus, we need only the 
coupling $C_{J/\psi \Lambda\, J/\psi \Lambda}$, which we take from Table~2 of Ref.~\cite{Feijoo:2022rxf}.

The main difference in the present study compared to that of \cite{Feijoo:2022rxf} is the method employed in the regularization of the loop function $G_l$ that enters in the BS equation in Eq.~(\ref{T_algebraic})), whose solution provides the unitarized scattering amplitude $t_{J/\psi \Lambda,J/\psi \Lambda}$ needed for the decay amplitude $\mathcal{M}$. Specifically, we now use the dimensional regulation method, already introduced in Eq.~(\ref{dim_reg}), instead of the original cutoff regulator ($q_{\rm max}$) employed in Ref.~\cite{Feijoo:2022rxf}. The reason lies in the fact that, when using a cutoff regulator and the on-shell momentum $q_{\rm on}$
is equal to $q_{\rm max}$, the real part of $G_l$ tends to infinity as the total energy $\sqrt{s}$ of the propagating $l-$channel increases, due to the non cancellation of the left branch of the principal value in the integral. The point is that the available invariant mass $m_{J/\psi \Lambda}$ is very large thereby favoring such a situation. On the contrary, the ${\rm Re}[G_l]$ obtained with dimensional regularization offers a smoother growth above the $l$th-threshold which allows one to circumvent the former drawback. The effects of implementing this new method are reflected in the location and width of the two states dynamically generated in this sector, as well as in the corresponding couplings of the channels considered to these states. The new properties of the poles are presented in Table~\ref{tab:spectroscopy}, where it can be appreciated that both states have been shifted towards higher energies, and the width for the one located at lower energies has increased notably and reduced for the second one when comparing to those of \cite{Feijoo:2022rxf}. In particular, the location of the $P_{\psi s}^\Lambda (4338)$ pentaquark, which is the one with significant changes, has been shifted by $75$ MeVs towards higher energies, but this state is not affecting much the invariant mass because of its small coupling to the $J/\psi \Lambda$ channel.
As a reminder, we would like to stress that, given the intrinsic spin-parity of the vector-meson and the one of the ground state baryons jointly with an interaction limited to s-wave, these states present a degeneracy in spin ($J^P=1/2-,3/2^-$).      

\begin{table}[h]
\caption{Pole content with $J^P=\frac{1}{2}^-,\frac{3}{2}^-$ for VLC Model (employing dimensional regularization) with their couplings  $g_i$ in the ($S=-1, \, I=0$) sector.} 
\smallskip
\centering
\begin{tabular}{c|c|c}
\hline
\hline 
  VLC Model &    $P_{cs}(4413)$   &  $P_{cs}(4570)$ \\
 \hline
$M\;\rm[MeV]$           & $4413.18$      & $4569.93$     \\
$\Gamma\;\rm[MeV]$      & $12.80$         & $8.32$   \\
\hline                         &   $ g_i$        &  $g_i$             \\
$J/\psi \Lambda$  &  $0.28-i0.05$          & $0.51-i0.02$      \\
$\bar{D}_s^*\Lambda_c$   &  $0.76+i0.29$   & $0.04-i0.02$     \\
$\bar{D}^*\Xi_c$  &  $2.99+i0.15$      & $-0.03+i0.02$      \\
$\bar{D}^*\Xi'_c$      &  $0.02-i0.05$  & $2.07-i0.13$     \\
\hline
\hline
\end{tabular}
\label{tab:spectroscopy}
\end{table}

It is important to clarify that the authors of Ref.~\cite{Feijoo:2022rxf} found a better candidate to associate with the $P_{\psi s}^\Lambda(4459)$, which couples mostly to the pseudoscalar-baryon $\bar{D}\Xi'_c$ channel. The present approach does not allow the coupling between the peudoscalar-baryon ($1/2^+$) and the vector-baryon ($1/2^+$) blocks, therefore precluding any trace of this state in the $m_{J/\psi \Lambda}$ invariant mass spectrum. However, in analogy with the three $P_{\psi}^N$ states observed in the $m_{J/\psi}p$ invariant mass spectrum in the $\Lambda_b$ decay~\cite{LHCb:2019kea}, the existence of an additional $P_{cs}$ state cannot be simply ruled out. It seems plausible that a more exhaustive analysis on the marked fluctuations present in the experimental $J/\psi \Lambda$ invariant mass can provide evidence of a new strange pentaquark. In any case, we should be cautious with the potential role of the $P_{cs}^\Lambda(4570)$ found in the VB ($1/2^+$) sector within the current approach. It is true that one might think that this dynamically generated state could be mimicking the observed $P_{\psi s}^\Lambda(4459)$, albeit with modified properties.

\subsection{Calculation of the invariant masses}
\label{subsec:dGam}
The double differential cross-section for the $\Xi^-_b \to J/\psi \Lambda K^-$ decay process reads:
\begin{equation}
\frac{d^2\Gamma}{dm^2_{K^- \Lambda}dm^2_{J/\psi \Lambda}}=
\frac{M_{\Xi_b}M_{\Lambda}}{64\pi^3M_{\Xi_b}^3}| \mathcal{M}(m_{K^- \Lambda},m_{J/\psi \Lambda})|^2   , 
\label{eq:double_diff_cross}
\end{equation}
where the amplitude $\mathcal{M}$ can be obtained from Eq.~(\ref{eq:Genamplitude2}) with the inputs provided by Eqs.~(\ref{eq:AmplscenarioA}), (\ref{eq:AmplscenarioB}) or (\ref{eq:AmplscenarioC}) depending on the scenario A, B or C, respectively.  

By fixing the invariant mass $m_{J/\psi \Lambda}$, we can integrate over $m_{K^- \Lambda}$ to obtain $d\Gamma/dm_{J/\psi \Lambda}$. In such a case, the limits for the $m^2_{K^- \Lambda}$ range are given by:
\begin{eqnarray}
(m_{K^- \Lambda}^2)_{\rm max} ={(E^*_{\Lambda}+E^*_{K^-})}^2 -{\Bigg(\sqrt{{E^*_{\Lambda}}^2-M^2_{\Lambda}}-\sqrt{{E^{*2}_{K^-}}-m^2_{K^-}}\Bigg)}^2   \nonumber \\ 
(m_{K^- \Lambda}^2)_{\rm min} ={(E^*_{\Lambda}+E^*_{K^-})}^2 -{\Bigg(\sqrt{{E^*_{\Lambda}}^2-M^2_{\Lambda}}+\sqrt{{E^{*2}_{K^-}}-m^2_{K^-}}\Bigg)}^2 , & \nonumber 
\end{eqnarray}
where the energies of $\Lambda$ and $K^-$ in the $J/\psi \Lambda$ rest frame read:
\begin{equation}
E^*_{\Lambda}=\frac{m_{J/\psi \Lambda}^2-m_{J/\psi}^2+M^2_{\Lambda}}{2m_{J/\psi \Lambda}} \, \, \,, \, \, \,
E^*_{K^-}=\frac{M_{\Xi_b}^2-m_{J/\psi \Lambda}^2-m^2_{K^-}}{2m_{J/\psi \Lambda}} . \nonumber
\end{equation}
Similar formulas can be derived when we instead fix the $m_{K^- \Lambda}$invariant mass and perform the integration over $m_{J/\psi \Lambda}$:
\begin{eqnarray}
\scalebox{0.95}{$( m_{J/\psi \Lambda}^2)_{\rm max}={(E^*_{\Lambda}+E^*_{J/\psi})}^2-{\Bigg(\sqrt{{E^*_{\Lambda}}^2-M^2_{\Lambda}}-\sqrt{{E^{*2}_{J/\psi}}-m^2_{J/\psi}}\Bigg)}^2$}   \nonumber \\
\scalebox{0.95}{$( m_{J/\psi \Lambda}^2)_{\rm min}={(E^*_{\Lambda}+E^*_{J/\psi})}^2-{\Bigg(\sqrt{{E^*_{\Lambda}}^2-M^2_{\Lambda}}+\sqrt{{E^{*2}_{J/\psi}}-m^2_{J/\psi}}\Bigg)}^2$} ,  \nonumber
\end{eqnarray}
with
\begin{equation}
E^*_{\Lambda}=\frac{m_{K^-\Lambda}^2-m_{K^-}^2+M^2_{\Lambda}}{2m_{K^- \Lambda}} \,\,\,,\,\,\,
E^*_{J/\psi}=\frac{M_{\Xi_b}^2-m_{K^- \Lambda}^2-m^2_{J/\psi}}{2m_{K^- \Lambda}} . \nonumber
\end{equation}

\subsection{Additional considerations for the theoretical invariant mass}
\label{subsec:normalization}

Finally, we address the discussion of how we normalize our predictions to be compared to the measured LHCb data and the assignation of values for the $\mathcal{N}$, $V_p$, $B$, $B'$ parameters entering the amplitude in Eq.~(\ref{eq:Genamplitude2}).\\
As already mentioned, the $V_p$ constant, preceding the dominant s-wave contribution to the decay amplitude, incorporates the combined strength of the weak and the hadronization processes. In general terms, one can fix $V_p=1$ assuming that its effective value will be transferred through a global $\mathcal{N}$ normalization of the theoretical invariant mass to the LHCb data. To reduce the amount of unknown quantities, $B$ and $B'$, which weight the different possible $J/\psi \Lambda$ spin-parity contributions to $K^-$ in p-wave, should take values relative to $V_p$. For simplicity, their corresponding values are taken to be similar ($B \approx B'$), and we assume a very moderate contribution of these p-wave terms to $\mathcal{M}$, namely around 10\% of the dominant $V_p$ term. Furthermore, given the dependence of the p-wave vertices on the momentum of the outgoing $K^-$ ($|\vec{q}|$), we add a multiplicative factor $10^{-3}$ to the previous 10\%, since the maximum $K^-$ momentum allowed by the phase space related to the present decay is $1054$ MeV. Thus, at the end, we estimate the values of $B$ and $B'$ to be $B=B'=\frac{1}{10}\times 10^{-3}$.\\
Finally, the overall normalization $\mathcal{N}$ is obtained by calculating the square root of the ratio between the integrated theoretical $K^-\Lambda$ invariant mass distribution over the experimental one in the $m_{K^-\Lambda}$ range going from $1680$ to $2700$ MeV. In the latter integration, the region slightly above the the $K^-\Lambda$ threshold has been dismissed given the large discrepancy between theory and experiment, as we will show in the next section.

\section{Results}\label{sec:results}
 Before comparing the results from the different models with the LHCb invariant mass spectra, we would like to first address in more detail the differences expected between the VBC and BCN models in the energy region probed by LHCb data on a simple theoretical basis.

The $\Xi^- _b$ decay mechanism studied in this work proceeds, besides the \LKMin, through the $K^- \Sigma^0$ ($E_{\rm thr.}$: 1686 MeV), $\bar{K}^0 \Sigma^-$ ($E_{\rm thr.}$: 1695 MeV), $\eta \Xi^-$ ($E_{\rm thr.}$: 1868 MeV) channels  via the corresponding $t_{i,K^- \Lambda}$ amplitudes (see diagram (b) in Fig.~\ref{fig:feynmandiag} and Eq.~(\ref{eq:amplitude1})). In principle, given the location of the threshold for these channels, above the \LKMin one, an enhancing effect on the corresponding $t_{i,K^- \Lambda}$ transitions in both models may be expected. However, the amount of such enhancement strongly depends on the strength of the transition amplitudes as well as on the couplings of the above mentioned channels to the \XRes state.\\
\begin{figure*}[t]
\begin{center}
\includegraphics[width=0.495\textwidth,keepaspectratio]{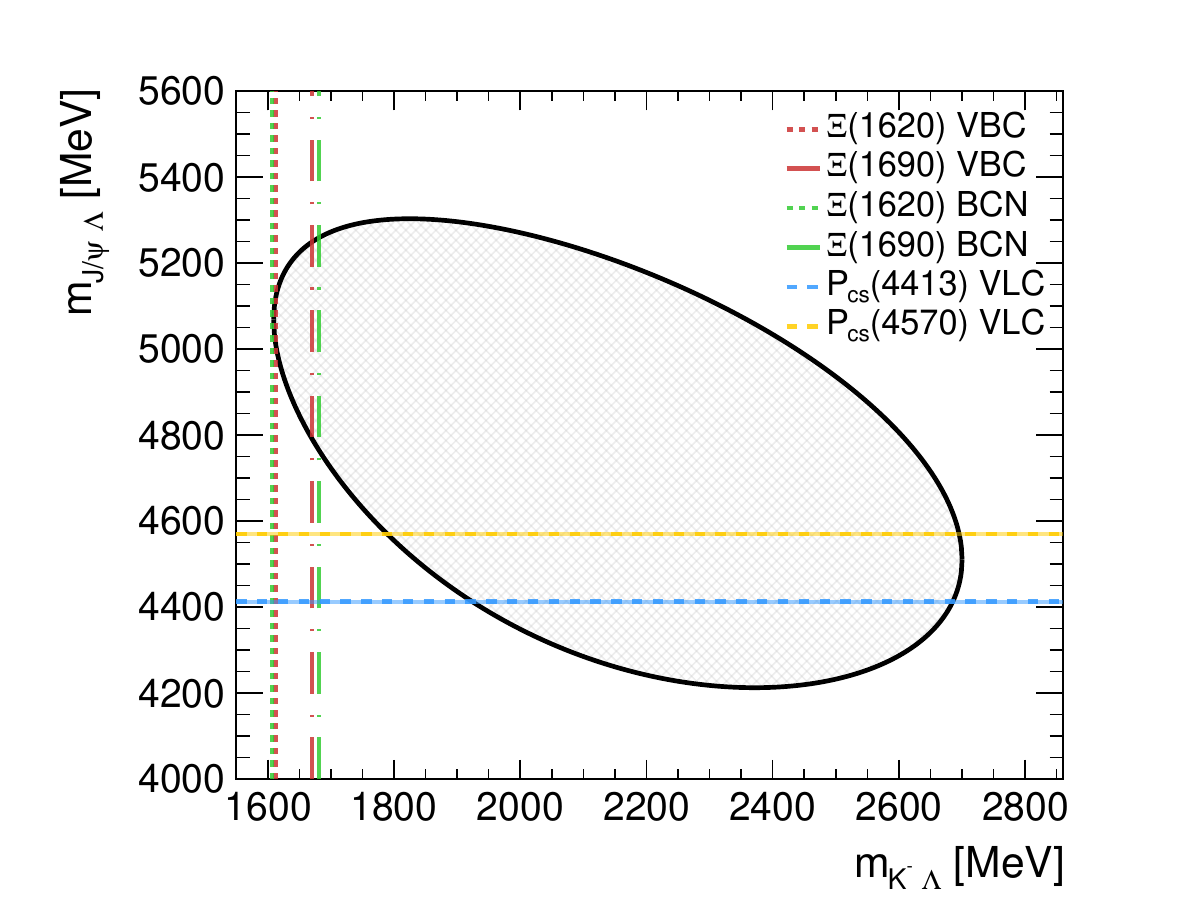}
\includegraphics[width=0.495\textwidth,keepaspectratio]{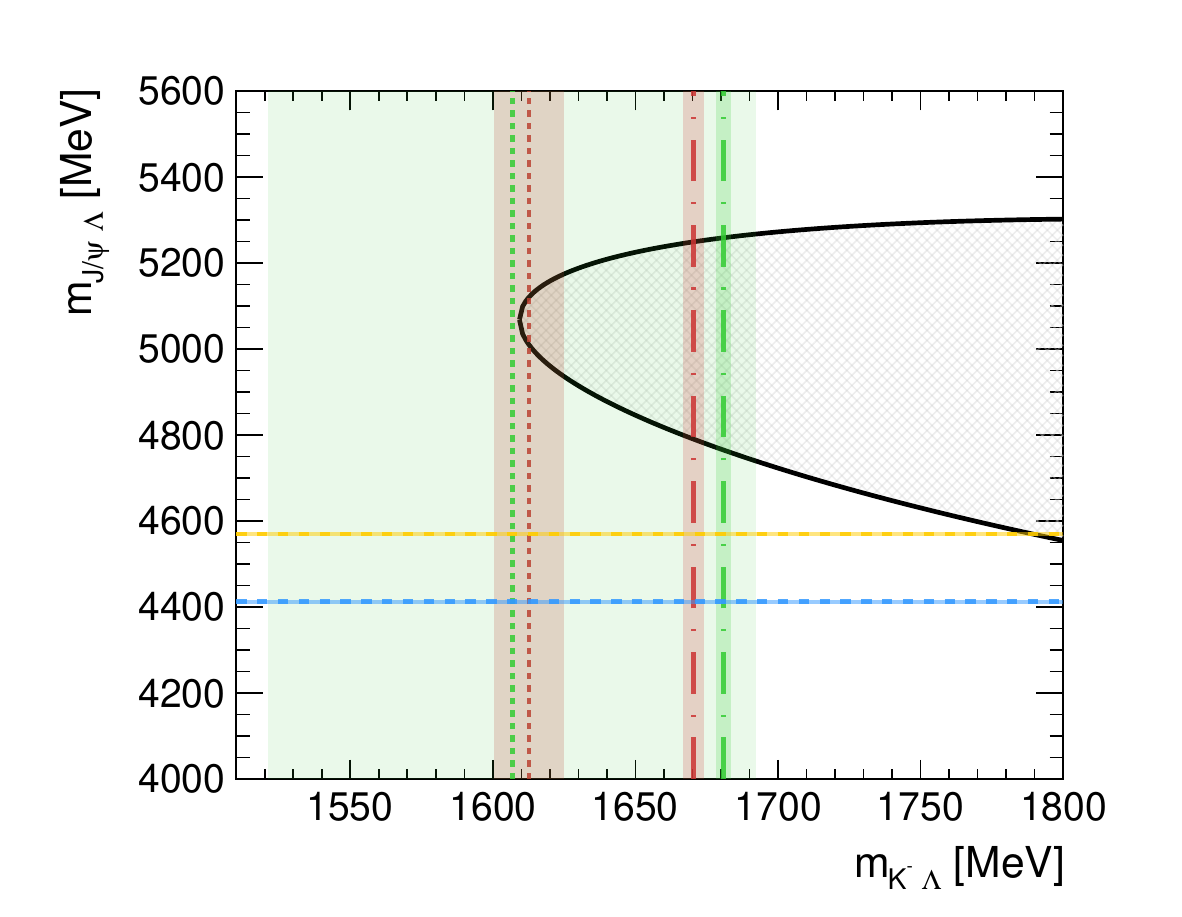}
\caption{(Color online). Right: Dalitz plot showing the allowed regions of $m_{J/\Psi \Lambda}$ (y-axis) and $m_{K^-\Lambda}$ (x-axis) invariant masses in the $\Xi_b \to J/\psi ~ K^- \Lambda $ decay. The different horizontal and vertical lines represent the dynamically generated states employing the theoretical models described in Sec.~\ref{subsec:FSI} (see the text for a detailed explanation). Left: zoom in the region close to the \LKMin threshold where the \XRes and \XResNovanta states are located. The filled area represents the widths obtained in VBC (red) and BCN (green) models for both poles. The enclosed hashed gray area represents the allowed kinematic region in the $\Xi^- _b$ decay.}
\label{fig:dalitz}
\end{center}
\end{figure*}
The VBC model has been constrained to the measured \LKMin CF~\cite{ALICE:2023wjz}, where the charged \XRes state was observed for the first time just above the \LKMin threshold with a reported mass of $\approx 1618$ MeV. 
In the obtained VBC parametrization of LECs and SCs, as can be seen in the supplemental material of~\cite{Sarti:2023wlg}, the above-mentioned amplitudes (particularly the $\bar{K}^0 \Sigma^-$, $\eta \Xi^-$ channels) not only play a relevant role in the reproduction of the measured $K^- \Lambda$ CF  but they are also crucial in the dynamical generation of the \XRes and \XResNovanta resonances. Given the location of the \XRes pole above the \LKMin threshold and the significant contributions from the $K^- \Sigma^0$, $\bar{K}^0 \Sigma^-$ and $\eta \Xi^-$ channels, we can expect the VBC model to provide a large signal in the \LKMin invariant mass distribution, particularly close to threshold.\\
The BCN model, on the contrary, only relies on very few experimental constraints, which makes feasible to constrain only the SCs in order to reproduce at best the  mass and width of the neutral \XRes and \XResNovanta states, provided by spectroscopic LHCb and Belle data. This leads to a completely different molecular interpretation of the \XRes with respect to the VBC results, more aligned with some previous U$\chi$PT studies limited to contact diagrams \cite{Ramos:2002xh,Garcia-Recio:2003ejq,Gamermann:2011mq,Sekihara:2015qqa}, that generate the \XRes below the $K^- \Lambda$ threshold with a sizeable width and coupling strongly to the $\pi \Xi$ channels. As a consequence, jointly with the fact that the transition $t_{\pi \Xi,K^- \Lambda}$ is not present in the decay amplitude given by Eq.~(\ref{eq:amplitude1}) since the $\pi \Xi$ pair is not produced in the primary hadronization ($h_{\pi \Xi}=0$), there would be a very diluted trace of the \XRes tail in the $K^- \Lambda$ invariant mass distribution from the BCN model. Additionally, since in the BCN molecular interpretation the \XRes couples to the $K^- \Lambda$ in a not negligible way, the Flatté effect also contributes to the reduction of the \XRes signal, as shown in Ref~\cite{Feijoo:2023wua}.\\

A quantitative example of the difference between the VBC and BCN pole properties is presented in Fig.~\ref{fig:dalitz} with the Dalitz plot for the three-body $\Xi_b \to J/\psi ~ K^- \Lambda $ decay process. The enclosed hashed area shows the kinematically allowed region. The mass of the poles for the \XRes (dotted) and \XResNovanta (dash-dotted) obtained in the BCN~\cite{Feijoo:2023wua} (green) and VBC~\cite{Sarti:2023wlg} (red) models are shown by the vertical lines.  The two dashed horitzontal lines represent the two dynamically generated pentaquark states in the VLC model~\cite{Feijoo:2022rxf} with the modified loop regularizaton.\\
On the right panel of Fig.~\ref{fig:dalitz}, we show a zoomed region of $m_{\LKMin}$ up to roughly 1800 MeV, focusing on the \XRes and \XResNovanta states predicted by the BCN and VBC models. The shaded areas represent the corresponding widths for each model.
As already shown in Table~\ref{tab:poles}, within the VBC parametrization, the \XRes pole is now located above the \LKMin threshold and it has a significantly smaller width compared to the \XRes pole from the BCN model, more in line with what is observed by ALICE in the \LKMin CF~\cite{ALICE:2023wjz} and by Belle in~\cite{BelleXi}. The \XResNovanta width becomes slightly larger in the VBC model and the mass is shifted by roughly $10$ MeV to lower values in contrast to that of Ref.~\cite{Feijoo:2023wua}.\\
The region of energies close to the \LKMin threshold showcases the strong impact of the \XRes and \XResNovanta on the $J/\psi \Lambda$ invariant mass spectrum within the range going from $4800$ to $5280$ MeV which can be appreciated in the right panel of Fig.~\ref{fig:dalitz}. The resonance properties, consequence of the inherent dynamics provided by each of the \LKMin amplitudes, as well as the combination of the different $t_{i,K^-\Lambda}$ contributions in the decay amplitude, tied to this particular decay mechanism, can provide valuable information about the goodness of the considered models when comparing to the experimental data.\\
\begin{figure}[ht]
\begin{center}
\includegraphics[width=0.5\textwidth,keepaspectratio]{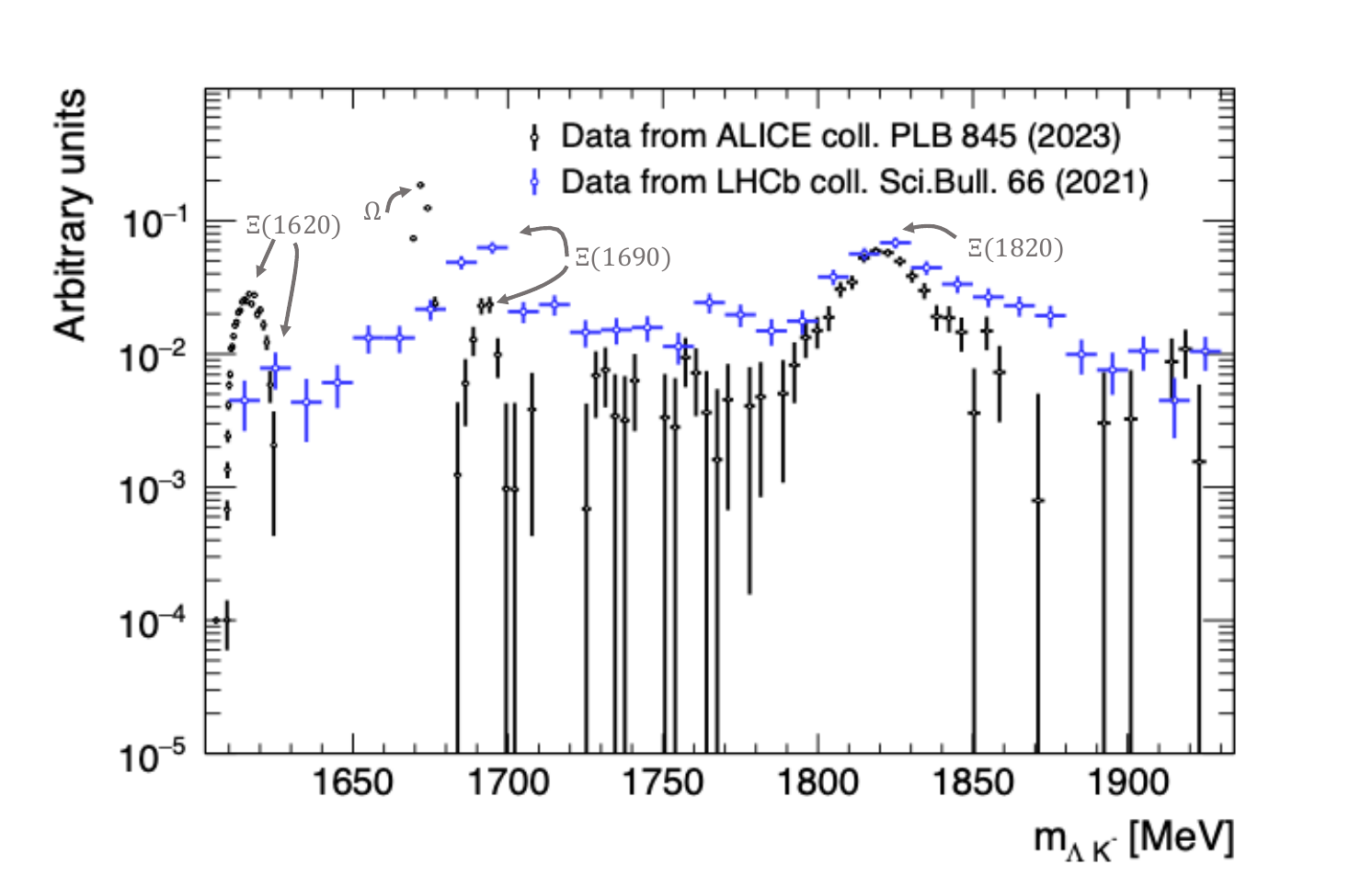}
\caption{(Color online). Invariant mass \LKMin~distributions obtained from ALICE correlation  \cite{ALICE:2023wjz} (black) and  LHCb $\Xi_b \to J/\psi ~ K^- \Lambda $ decay \cite{LHCb:2020jpq} (blue) data. For comparison purposes, both data have been normalized within the published range of energies. Due to the logarithmic scale in the y-axis, the negative values in the ALICE data are not shown.}
\label{fig:ComparisonIMLHCbALICE}
\end{center}
\end{figure}
Moreover, by inspecting along the $m_{\LKMin}$ axis in the energy range from $1800$ MeV onwards, one can expect an important contribution from diagram (c) of Fig.~\ref{fig:feynmandiag} in the  $K^- \Lambda$ invariant mass distribution, stemming from the dynamically generated $P_{cs}$ in the VLC model after integrating over $m_{J/\psi \Lambda}$. This will be clearly seen when comparing among the different scenarios (A, B and C) where we deal with different configurations of spin-parities and speculate with the potential existence of a third pentaquark, in analogy with the observed $P_\Psi^N(4312), P_\Psi^N(4440)$ and $P_\Psi^N(4457)$
states.

In Fig.~\ref{fig:ComparisonIMLHCbALICE} we show the comparison between the published data points on the \LKMin invariant mass spectra delivered by ALICE in~\cite{ALICE:2023wjz} (black) and by LHCb in~\cite{LHCb:2020jpq}. The invariant mass distribution reported by ALICE has been obtained from the difference between the same- and mixed-event distributions used to build the \LKMin correlation function and, analogously to the CF, it exhibits the presence of the charged \XRes state around $1618$ MeV. Preliminary results on the opposite charge $\pi\Xi$ CF measured by ALICE~\cite{SQMXiPiALICE} also show the presence of the neutral \XRes which seem to confirm the location of this state previously observed by Belle~\cite{BelleXi} in the same channel, around $\approx 1610$ MeV.\\
Continuing the parallelism between correlation and spectroscopic data, we focus now on the \LKMin LHCb data. In the vicinity of the \LKMin threshold, a hint of a structure can be seen in the invariant mass measured by LHCb, which however is currently not significant to claim the presence of the \XRes state in this data sample. Such a comparison must be taken with caution since the \lmb and the kaons entering in the two measurements have been obtained with different kinematic and topological cuts, which for example lead to the $\Omega \rightarrow \lmb K^-$ peak clearly visible in the ALICE data. Note that the spectrum reported by ALICE in~\cite{ALICE:2023wjz} is not corrected for acceptance and efficiency effects, typically taken into account in spectroscopy measurements. However, since the particles entering the low momentum region of a two-body CF measurement have similar momenta, we do not expect such effects to significantly modify the ALICE spectrum.
Despite such differences and a reduced \LKMin statistics in the LCHb data due to the constraints to the $\Xi_b$ decay selection, we can observe that the peak locations of the higher mass $\Xi^\ast$ resonances, the \XResNovanta and $\Xi$(1820), are  in overall agreement between the two datasets. The deviation among the two measurements is confined to the region below 1670 MeV. 
The high-precision achieved with ALICE data allows to clearly see the \XRes peak just above the \LKMin threshold while a much weaker signal is present in the LHCb spectra.
The reason behind such a more enhanced signal of the \XRes pole in the ALICE invariant mass and related CF observable with respect to the LHCb spectroscopic measurements should, beyond any possible differences at the level of data-sample statistics and particle selection, be found in the different production mechanism leading to the formation of a molecular state such as the \XRes that can be accessed and probed with femtoscopy or with spectroscopy data.\\

As mentioned at the beginning of this section, the $\Xi^- _b$ decay is mainly probing the $S=-2$ meson-baryon FSI via the \LKMin, $\eta\Xi$ and $\bar{K}\Sigma$ channels, with the latter being the dominant contribution thanks to the larger transition weights ($h_{K^- \Sigma^0}=\frac{1}{\sqrt{2}}$, $h_{\bar{K}^0 \Sigma^-}=1$). In correlation measurements, the transition amplitudes  $t_{i,\LKMin}$ also play a crucial role, by entering explicitly in the evaluation of the wave-function $\psi_{i,\LKMin}$ needed to determine the inelastic contributions to the CF (see~\cite{Sarti:2023wlg,Haidenbauer:Coupled,ALICE:CoupledKp,Vidana:2023olz,Albaladejo:2023pzq, Feijoo:2024bvn} for details). Each of these contributions are weighted by the so-called production weights $\omega^{\rm prod.} _i$ which are related to the amount of initially produced $i =  \PiMinXiZ, \PiZXiMin, \KMinSigZ, \KZbarSigMin, \EtaXiMin$ pairs. In 
the \LKMin  CF, the $\pi\Xi \rightarrow \LKMin$ contribution carried the largest weight, while the $\eta\Xi$ production was significantly reduced. Nevertheless, the LECs and SCs parameters fitted to the CF delivers a VBC model in which both \XRes and \XResNovanta poles show a significant coupling to \EtaXiMin and $\bar{K}\Sigma$ channels and present properties more in line with the available measurements~\cite{BelleXi,LHCb:2020jpq,SQMXiPiALICE}.\\
This is just a first example of how, at the theoretical level, the CF and invariant mass observables can in principle be sensitive to different contributions of the underlying dynamics. Dynamics which, as in the meson-baryon $S=-2$ case at hand, presents a rich coupled-channel structure whose full understanding would require to have at our disposal experimental constraints on as many channels as possible. The experimental access to various of the pairs in this sector is challenging since we are dealing with strange baryons and mesons. Hence, we would like to stress the importance of starting to consider differential data, such as correlations and spectroscopy, which can be able to provide constraints either on the same or on several different channels within the same sector. In this first work, we aim at showing a practical example of such approach by exploring how a model that has been constrained to a \LKMin correlation measurement compares with spectroscopic \LKMin data.

In the following, we proceed to discuss in detail the comparison between the LHCb data and the models for the \LKMin and \JPsiL interactions described in Sec.~\ref{subsec:FSI}.
\noindent
\begin{figure*}[t]
\begin{center}
\includegraphics[width=0.49\textwidth,keepaspectratio]{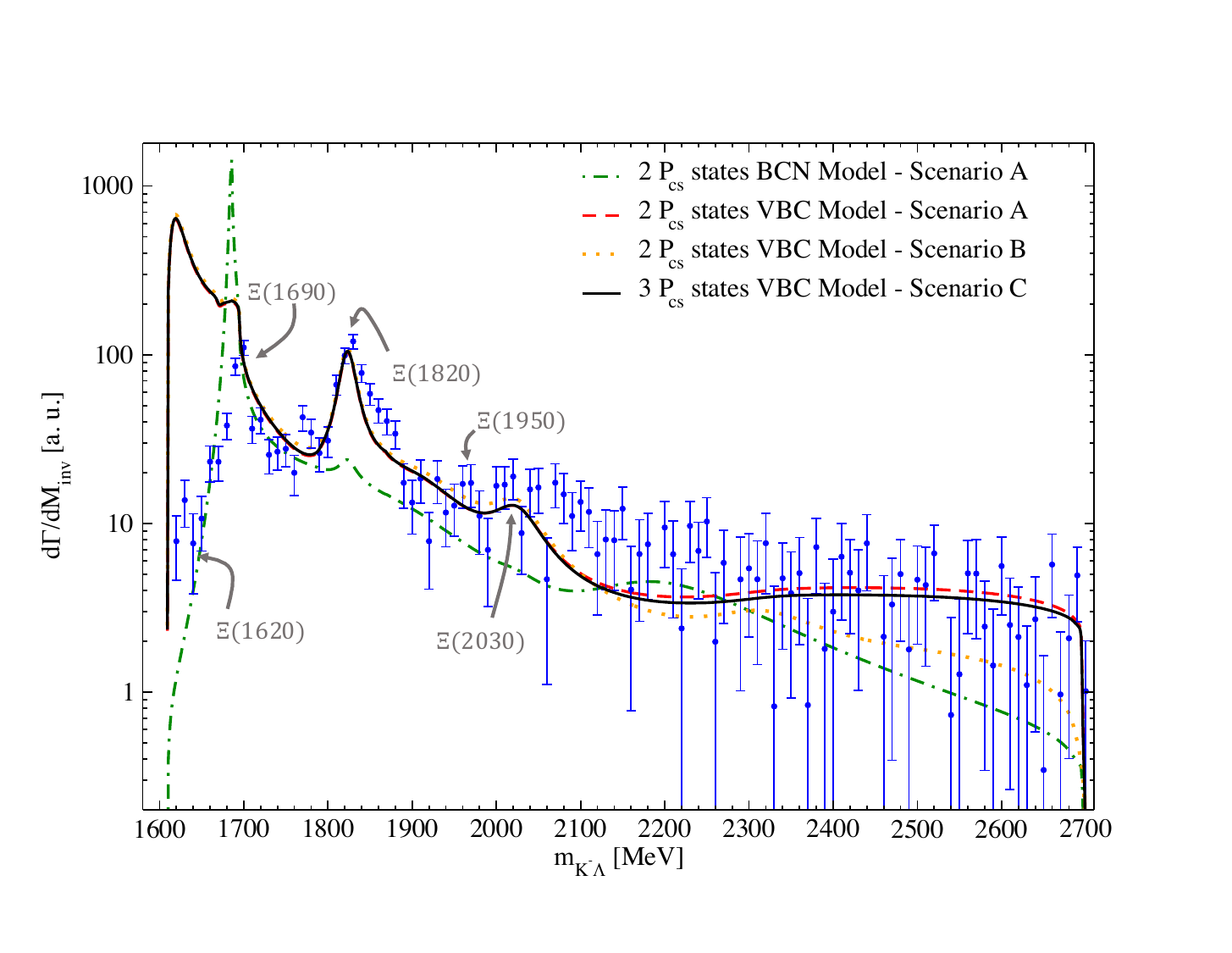}
\includegraphics[width=0.49\textwidth,keepaspectratio]{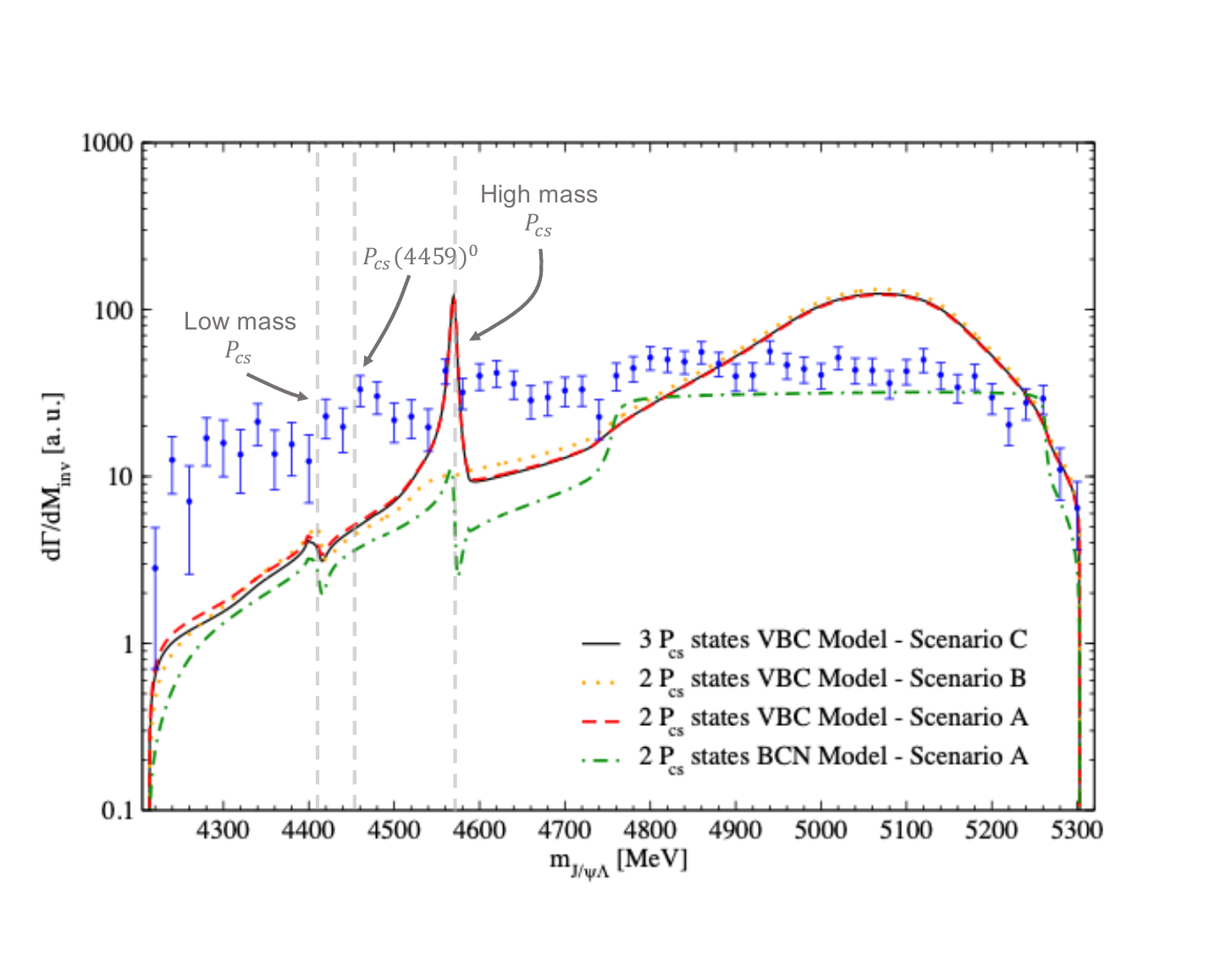}
\caption{(Color online). \LKMin (left) and \JPsiL (right) invariant mass spectra. The blue markers represent the measured data from the LHCb collaboration reported in~\cite{LHCb:2020jpq}, normalized to the theoretical calculations. The lines represent the theoretical calculations assuming either the VBC or BCN \LKMin amplitude along with the different $P_{cs}$ scenarios.}
\label{fig:InvMassLines}
\end{center}
\end{figure*}
In Fig.~\ref{fig:InvMassLines}, we show the calculated invariant mass spectra of \LKMin (left) and \JPsiL (right) pairs compared to the measured LHCb data. First, assuming the scenario A on the two $P_{cs}$ spin configurations, we compare the calculations obtained within the VBC (red dashed) and BCN (green dot-dashed) models. It can be seen that the VBC results show a much better agreement with the data in the $m_{\LKMin}$ region above $1680$ MeV. This is an interesting output because, despite the inclusion of the three heavy $\Xi^\ast$ states is done similarly in both models via the BW-type recipes, and both models have the same $T_{1/2} ^{\rm{p-wave}}$ and $T_{3/2} ^{\rm{p-wave}}$ terms in Eq.~(\ref{eq:Genamplitude2}), the BCN parametrization fails at reproducing the data in the resonant region between $1800$ and $2250$ MeV and in the high energy one. This reinforces the relevance of having high-precision data, given by the measured \LKMin CF in this case, to constrain the VBC scattering matrix elements, whose interplay with the BW contributions delivers a good description above $1680$ MeV. A plausible explanation of such agreement can be found in the preliminary discussion at the beginning of this section, where we advanced the potential role of the VBC constrained $t_{i,K^- \Lambda}$ ($i=K^- \Sigma^0, \bar{K}^0 \Sigma^-,\eta \Xi^-$) because of their direct contribution to the $\Xi^- _b$ decay amplitude.\\
The energy region close to the \LKMin threshold requires a more detailed discussion. The BCN model, delivering the \XRes pole below the \LKMin threshold (see Table~\ref{tab:poles}), is indeed not able to reproduce the experimental enhancement that could well be associated to the \XRes resonance, while the VBC model overshoots the experimental structure by almost two orders of magnitude because of the pole found in the ($S=-2$) scattering matrix at $1613$ MeV (see Table~\ref{tab:poles}). The BCN model behaves as expected, meaning that any trace of the \XRes vanishes for the reasons advocated in the preliminary considerations. In contrast, the large strength provided by the VBC model can be understood by the strong coupling that the dynamically generated \XRes state has to the $K^- \Lambda$ and $\eta \Xi^-$ channels (check Table~II in Ref~\cite{Sarti:2023wlg}), whose related $t_{K^- \Lambda,K^- \Lambda}$ and $t_{\eta \Xi^-,K^- \Lambda}$ play an active role in the decay amplitude $\mathcal{M}$.\\
At slightly higher energies, where the \XResNovanta state should appear, both models overestimate the data, about a factor $2$ for the VBC model and approximately one order of magnitude for the BCN one. However, it has to be reminded that the VBC calculations reported in Fig.~\ref{fig:InvMassLines} are obtained by taking only the central values of the fitted LECs and SCs in Table~I of Ref~\cite{Sarti:2023wlg}. A more detailed comparison between data and the VBC model with the inclusion of the uncertainties on its parameters will be discussed below in Fig.~\ref{fig:InvMassSigmas}.\\
If we now consider only the VBC model and investigate the sensitivity of the \LKMin invariant mass spectra to the different $P_{cs}$ assumptions, we see that we cannot discriminate among the three scenarios A, B and C. However, it seems that the experimental trend at higher energies is better described by those scenarios which incorporate a high mass $P_{cs}$ with $J^P=1/2^-$, i. e. scenario A (dashed line) and scenario C (solid line).
\begin{figure*}[t]
\begin{center}
\includegraphics[width=0.495\textwidth,keepaspectratio]{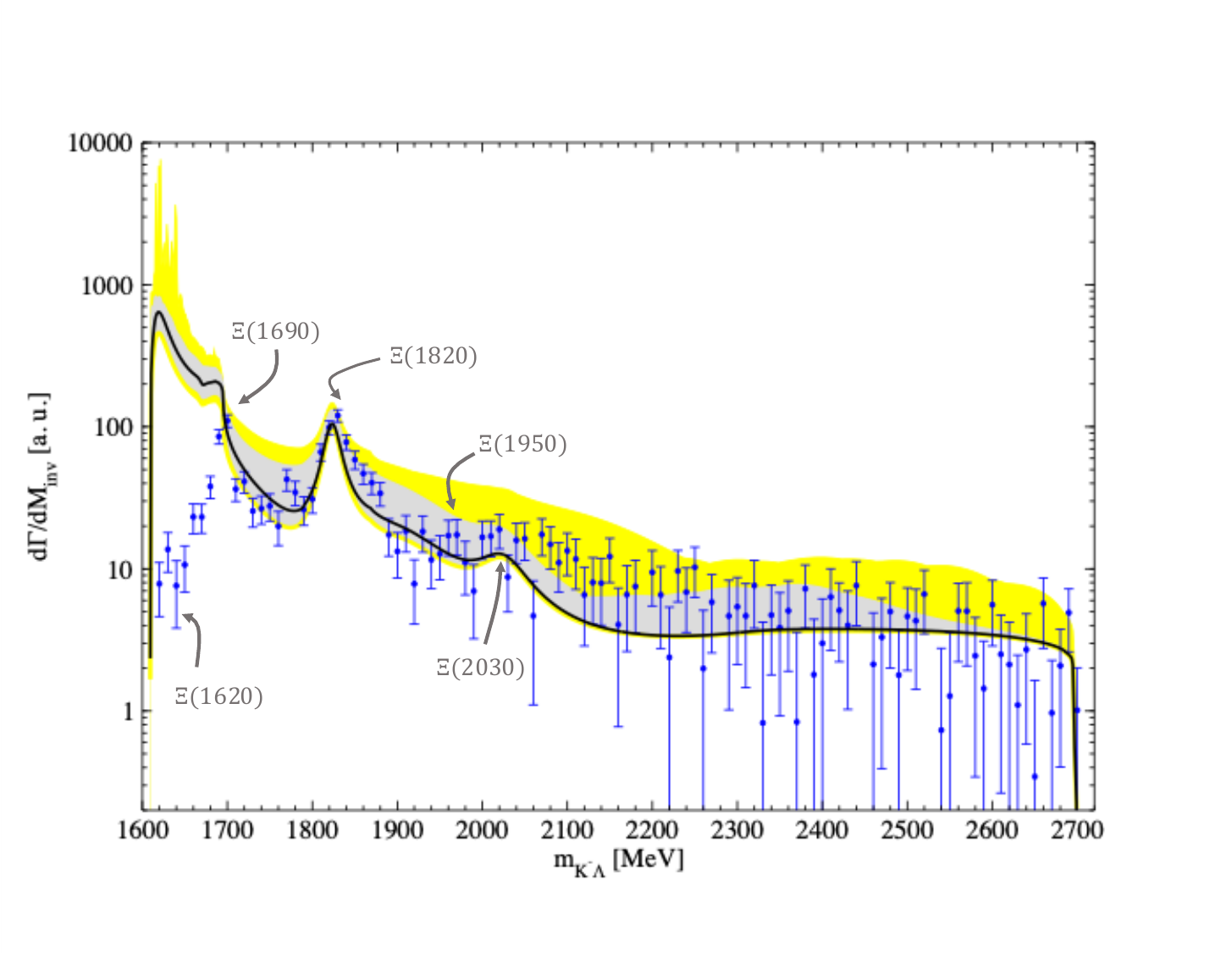}
\includegraphics[width=0.495\textwidth,keepaspectratio]{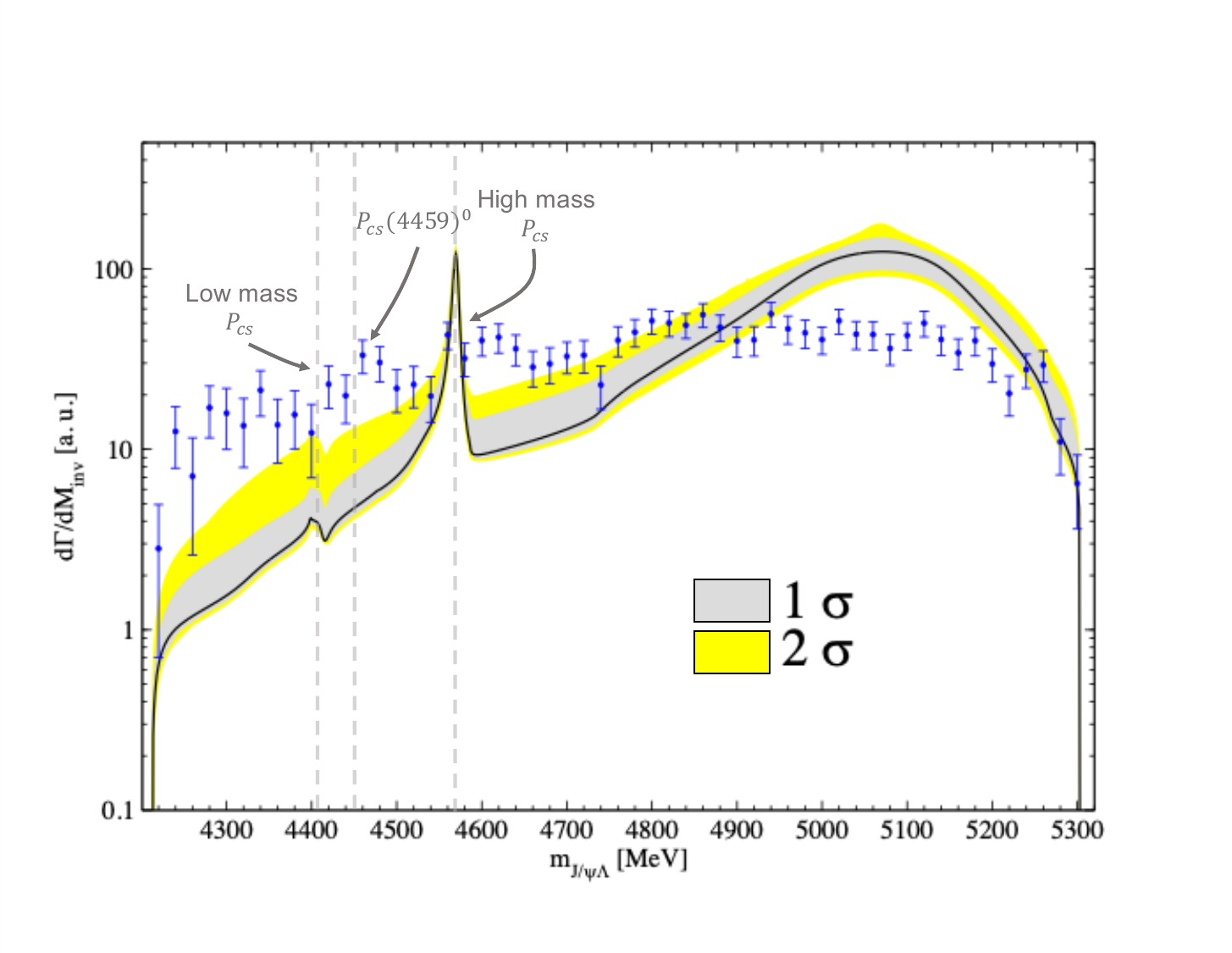}
\caption{(Color online). \LKMin (left) and \JPsiL (right) invariant mass spectra compared to the VBC + Scenario C model. The blue markers represent the measured data from the LHCb collaboration reported in~\cite{LHCb:2020jpq}, normalized to the theoretical calculations. The lines represents the theoretical calculations assuming the center values of the extracted LECs and SCs for the VBC model in Ref.~\cite{Sarti:2023wlg}. The grey and yellow bands shows the theoretical predictions when we include one or two standard deviations as uncertainties on the LECs and SCs (see Table ~II in Ref.~\cite{Sarti:2023wlg}).}
\label{fig:InvMassSigmas}
\end{center}
\end{figure*}
For completeness, and bearing in mind that this is out of the scope of the present study, we also display the corresponding \JPsiL invariant mass obtained for the different models and scenarios. It is convenient to remind that we deviated from the LHCb assumption of one single (or a double peak structure) $P_{cs}(4459)^0$ pentaquark state and we tested the presence of at least two states (see Table~\ref{tab:spectroscopy}) with different $J^P$ according to the findings of \cite{Feijoo:2022rxf}. As already mentioned, the different pole properties of the $\Xi^\ast$ due to differences in the $S=-2$ meson-baryon interaction obtained from the VBC and BCN scattering amplitudes have an impact on the \JPsiL invariant mass. This can be clearly seen on the right panel of Fig.~\ref{fig:InvMassLines}. The calculations based on the BCN or VBC parametrizations are overall not able to reproduce the data in the whole energy region. A striking difference can be observed in the strength of the high mass $P_{cs}$ state and at energies above it.
The reason lies in the fact that within the VBC model a better description of the \LKMin region around the $\Xi$(1820) is achieved, which corresponds, as shown in the Dalitz plot in Fig.~\ref{fig:dalitz}, to the location of the \JPsiL pole at $\approx 4570$ MeV. More precisely, the interference pattern generated between the different terms in the decay amplitude seems to favor the VBC model around this region. The enhancement for $m_{J/\Psi \Lambda}$ above $4800$ MeV provided by the VBC model, independently from the $P_{cs}$ scenarios, stems from the influence of the different locations and widths of the \XRes and \XResNovanta states (see right panel of Fig.~\ref{fig:dalitz}). This is because in the VBC approach, both states enter having sizable couplings (and much smaller \XRes width) to $K^- \Lambda$, $K^- \Sigma^0$, $\bar{K}^0 \Sigma^-$, $\eta \Xi^-$ whose corresponding $t_{i,K^- \Lambda}$ transitions have proven to be important in the $K^- \Lambda$ CF and are explicitly present in the decay amplitude $\mathcal{M}$. This is in contrast to the BCN model behavior that, in the high energy region, reaches a plateau due to the large width obtained for the \XRes resonance which dilutes any residual structure of such a state. The shallow slope in the plateau can be caused by the presence of the \XResNovanta resonance through the $t_{\bar{K} \Sigma,K^- \Lambda}$ amplitudes. As final remark on this part of the results, we would like to stress that the comparison between the LHCb \JPsiL invariant mass data and the assumed \LKMin modeling shows sensitivity in a limited energy region and it was performed in order to provide a complete investigation of all the available data provided in~\cite{LHCb:2020jpq}. A better description of the \JPsiL invariant mass is beyond the scope of this work.\\

To conclude this section we present the comparison between data and a model including the VBC parametrization, with the corresponding uncertainties on the LECs and SCs extracted from the fit to the measured \LKMin CF in~\cite{Sarti:2023wlg}. We consider only the scenario C with three $P_{cs}$ states for the \JPsiL interaction as an illustrative case, since no significant differences with respect to A and B were observed.
In Fig.~\ref{fig:InvMassSigmas} we show the results for the \LKMin (left) and \JPsiL (right) invariant mass distribution considering the 1$\sigma$ (grey band) and 2$\sigma$ (yellow band) variation variation on the LECs and SCs parametrization of the VBC model (reported in Table I in~\cite{Sarti:2023wlg}).
Within this quoted uncertainty ranges and following a similar procedure as in~\cite{Sarti:2023wlg}, we create the bands by generating thousands of theoretical spectra from randomly sampling each value of LEC and SC within the 1 or 2 $\sigma$ interval. For each value of energy, the distribution of the produced decay rates was checked and found to be quite homogeneous, thus the error band has been calculated assuming a uniform distribution.\\
Both invariant mass data can be well reproduced within the 1$\sigma$ band in the intermediate and high energy region. The larger discrepancy between model and data occurs close to the \LKMin threshold, up to the \XResNovanta, where the \XRes pole is located. Correspondingly, such differences in the \LKMin affect as well the agreement to the measured \JPsiL spectra above 4900 MeV, as shown in Fig.~\ref{fig:dalitz}.\\
The enhancement in the VBC prediction stems from the sensitivity of the underlying amplitude to the \XRes pole generation, and hence to the values assumed for the LECs and SCs. The work of Ref.~\cite{Sarti:2023wlg} demonstrated that these parameters can be constrained using CF data and provided a robust procedure to that end. What was also shown explicitly by the uncertainties reported on the LECs, which regulate the strength of the transition potentials $V_{ij}$, is that more high precision data in the couplings of other channels to the \LKMin pair are needed and in particular in the region of the \XRes and \XResNovanta states. As mentioned above, these kind of data should become soon available thanks to preliminary results on the $\pi^+\Xi^-$ CF recently reported by the ALICE collaboration~\cite{SQMXiPiALICE}, where both the \XRes and \XResNovanta states can be seen in the data. Hence, once available, we plan to perform future analyses aiming at further constraining the $S=-2$ meson-baryon amplitude in a wide energy region, from the $\pi\Xi$ threshold up to the vicinity of the $\eta\Xi$ one.



\section{Summary and Conclusion}
\label{sec:conclusions}

In this work, we presented a first study on the ability of a U$\chi$PT model at NLO (VBC), constrained to the recent $K^-\Lambda$ correlation data measured by ALICE, to describe the $K^-\Lambda$ invariant mass data obtained by the LHCb collaboration in the $\Xi^-_b \to J/\psi \Lambda K^-$ decay. We explored the interplay between the \LKMin and \JPsiL interaction through the $\Xi_b^-$ decay amplitude associated to an internal emission mechanism.

The VBC model features a novel molecular interpretation of the two \XRes and \XResNovanta molecular states, which is more in line with the available measured properties and rather different from previous models in the same $S=-2$ sector constrained to $\bar{K}$N data. The \LKMin mass spectrum, within the same U$\chi$PT framework at NLO, is calculated for a model fitted to $S=-1$ data (BCN) and for the VBC model. The latter provides a better description of the measured \LKMin mass distribution in the range of energies above the \XResNovanta state. 

The \JPsiL scattering amplitude is obtained within a unitary extension of a chirally motivated interaction based on HQSS and LHG which dinamically generates $2$ degenerate $P_{cs}$ states.

 We probed the sensitivity of both the \LKMin and \JPsiL invariant mass spectra to different assumptions on the spin-parity and multiplicity of the generated $P_{cs}$ states. Our results show a non-relevant effect of the different pentaquark scenarios on both \LKMin and \JPsiL spectra. Slightly larger differences are observed within the two \LKMin models in the \JPsiL system, however both models show an overall disagreement with the LHCb data.\\
The larger discrepancy between the VBC and BCN parametrization is visible in
the energy region close to the \LKMin threshold, dominated by the presence of the \XRes state, which affects in a different way both the measurements and the theoretical predictions. Such tension with the measured \LKMin spectroscopic data indicates the importance of pinning down precisely the properties and composition of the \XRes state, which necessarily requires the need for additional experimental input on the other channels coupling to the \LKMin system. High-precision constraints on channels such as \XiPi and $\bar{K}\Sigma$, up to the $\eta\Xi$ energy region could allow to better constrain the parameters of the U$\chi$PT interaction describing the MB $S=-2$ sector. We aim at performing future studies in this direction by exploiting preliminary correlation data on \XiPi pairs recently presented by the ALICE collaboration and additional femtoscopy measurements involving charged $\Sigma$ baryons which will become available in the next years within the on-going LHC Run 3 and future Run 4 data-taking.


\section{Acknowledgements}
V. M. S. and A. F. are very grateful to Dr.\,\,Bo Fang for providing us the published LHCb experimental data. A. F. was supported by ORIGINS cluster DFG under Germany’s Excellence Strategy-EXC2094 - 390783311 and the DFG through the Grant SFB 1258 ``Neutrinos and Dark Matter in Astro and Particle Physics”. V. M. S. was supported by the Deutsche Forschungsgemeinschaft (DFG) through the grant MA $8660/1-1$. J. N. was supported by the Spanish Ministerio de Ciencia e Innovaci\'on (MICINN) and European FEDER funds under Contracts No.\,PID2020-112777GB-I00, PID2023-147458NB-C21 and CEX2023-001292-S (Unidad de Excelencia “Severo  Ochoa”); by Generalitat Valenciana under contract CIPROM/2023/59. A. R. has been supported by MCIN/ AEI/10.13039/501100011033/ and by FEDER UE through grant PID2023-147112NB-C21 and through the “Unit of Excellence María de Maeztu 2020-2023” award to the Institute of Cosmos Sciences, grant CEX2019-000918-M as well as by the EU STRONG-2020 project.

\bibliographystyle{apsrev4-1}
\bibliography{refs.bib}

\end{document}